\documentclass[12pt]{wlscirep}
\usepackage[utf8]{inputenc}
\usepackage[T1]{fontenc}
\usepackage{tabularx,subcaption,fancyhdr,mathtools}
\usepackage{xcolor}
\newcommand{\tcb}[1]{\textcolor{blue}{#1}}

\usepackage{multirow}

\title{Solving Navier-Stokes Equations Using Data-free Physics-Informed Neural Networks With Hard Boundary Conditions
\vspace{+1cm}
}

\author[1,2]{Ritik Pal}
\author[1]{Soubhik Mukherjee}
\author[3]{Urmi Dutta}
\author[1]{Arghya Choudhury$^{}$}

\affil[1]{Department of Physics, Indian Institute of Technology Patna, 
Bihar, 801106, India}
\affil[2]{Indian Institute of Science Education and Research Thiruvananthapuram, Vithura, Kerala, 695 551, India}
\affil[3]{Department of Geology, Patna University, Patna, Bihar, 800005, India}

\affil[*]{Email: \tcb{ritik24@iisertvm.ac.in;  soubhik\_2321ph13@iitp.ac.in;  urmidutta@pup.ac.in; arghya@iitp.ac.in}}


\begin{abstract}
In recent years, Physics-Informed Neural Networks (PINNs) have 
emerged as a powerful and robust framework for solving nonlinear differential equations across a wide range of scientific and engineering disciplines, including biology, geophysics, astrophysics and fluid dynamics. In the PINN framework, the governing partial differential equations, along with initial and boundary conditions, are encoded directly into the loss function, enabling the network to learn solutions that are consistent with the underlying physics. In this work, we employ the PINN framework to solve the dimensionless Navier-Stokes equations for three two-dimensional incompressible, steady, laminar flow problems without using any labeled data. The boundary and initial conditions are enforced in a hard manner, ensuring they are satisfied exactly rather than penalized during training. We validate the PINN predicted velocity profiles, drag coefficients and pressure profiles against the conventional computational fluid dynamics (CFD) simulations for moderate to high values of Reynolds number ($Re$). It is observed that the PINN predictions show good agreement with the CFD results at lower $Re$.  
We also extend our analysis to a transient condition and find that our method is equally capable of simulating complex time-dependent flow dynamics.
To quantitatively assess the accuracy, we compute the $L_2$ normalized error, which lies in the range $\mathcal{O}(10^{-4})$ - $\mathcal{O}(10^{-1})$ for our chosen case studies.
\end{abstract}

\begin{document}

\flushbottom
\maketitle
%
%
\thispagestyle{empty}

\tableofcontents

\section{Introduction}

\label{sec:intro}

Over the past few decades, the advancements in computational fluid dynamics (CFD) across various industrial and academic research areas, such as turbulent flows, Fluid-Structure Interaction (FSI), heat transfer, multiphase flow, etc., have raised the demand to invent more precise, efficient and cost-effective simulations. Disciplines like biomedical engineering, automotive design, and geophysics deal with real-world fluid systems governed by some Partial Differential Equations (PDEs), which can be numerically solved using conventional 
finite difference, finite element, finite volume or spectral methods~\cite{temam2024navier, BROOKS1982199, fem1, fvm, GK_CFDbook,DUTTA20141,dutta2016role}.
Along with the continuity equation, the fundamental dynamics of fluid flow are described by the Navier-Stokes equations (NSEs), which are highly nonlinear and represent the momentum balance of the system. 
However, solving complex practical applications can be computationally very expensive and challenging. 
The examples can be systems with highly complex geometries, extreme values of parameters (very large Reynolds number, Re), minimal knowledge about the underlying physics or boundary conditions (BCs), and complicated meshing (moving boundary problems). Traditional CFD methods are limited in their ability to handle inverse problems. However, by using the observed flow data along with modern machine learning (ML) algorithms, one can predict the unknown input parameters.

Numerous fields of science and engineering have greatly benefited from the use of machine learning, typically for classification and regression problems. In recent years, researchers have started to solve PDEs or fluid dynamics problems via a novel approach known as Physics Informed Neural Networks (PINNs), which uses Neural Networks (NN) and Deep Learning (DL) algorithms. First introduced by Raissi et al. in 2019 \cite{RAISSI2019686}, PINNs integrate the governing differential equations of a physical system 
 by incorporating them as part of the loss function for the neural network. 
 This innovative approach leverages the use of automatic differentiation (AD) 
\cite{baydin2018automatic}, a powerful feature provided by modern DL frameworks like TensorFlow \cite{abadi2016tensorflow} and PyTorch \cite{paszke2019pytorch}. AD \cite{baydin2018automatic} tracks each and every operation performed on the input variables, and using the chain rule; it computes the derivatives of the output with respect to the input variables. The derivatives in the PDEs are evaluated using AD.
Thus, PINN provides an additional advantage over conventional CFD techniques by bypassing the mesh generation.

 PINNs have been successfully employed to solve a variety of linear and nonlinear PDEs, both in data-driven and physics-driven applications, e.g., the study of vortex-induced vibration~\cite{raissi2019deep}, analyzing inverse flow problems using hidden fluid mechanics (HFM) framework~\cite{raissi2020hidden},
estimation of the unknown input parameters and the constitutive relationship for flows in porous media~\cite{tartakovsky2018learning}, formulation of \texttt{NSFnets} to study laminar and turbulent flows in 3D~\cite{jin2021nsfnets}, training of Deep Neural Network (DNN) using multi fidelity data~\cite{meng2020composite}, exploring the missing flow dynamics with limited amount of data~\cite{xu2021explore}. One of the key strengths of PINNs lies in their ability to provide data-driven solutions to PDEs. By incorporating the PDE as a term in the loss function, PINNs can learn the underlying physics from noisy or scarce data \cite{satyadharma2024assessing}, enabling them to capture the physics-based trends and accurately extrapolate solutions to regions where data is not available~\cite{meng2020composite, xu2021explore,satyadharma2024assessing}. Additionally, PINNs can be used to solve the inverse problem, where the known solution is utilized to infer or estimate the unknown underlying parameters or hidden dynamics of the governing PDEs~\cite{raissi2020hidden, tartakovsky2018learning}. However, PINNs can be computationally expensive, and the convergence of the network can be slow. Due to this, hyperparameter optimization becomes even more challenging. Also, the convergence of the solution is highly dependent upon the complexity of the problem and is not always guaranteed. Several modifications have been proposed in the literature to address these problems, accelerate the convergence and improve the performance of the network~\cite{krishnapriyan2021characterizing, basir2022critical, wang2022and, jagtap2020locally,wang2020understanding}. For more detailed analyses of the potential applications of the methods and tools of PIML and its algorithms, frameworks or limitations, see refs~\cite{karniadakis2021physics, cuomo2022scientific,en16052343} and references therein. From these works, it can be noted that data-driven PINN applications always require a certain amount of labeled simulation or experimental data to train the DNN or solve the PDEs. However, several groups have implemented unsupervised learning within the PINN framework without incorporating any data points \cite{sun2020surrogate, biswas2023three, shah2024physics}. 
Sun et al. showed that the implementation of initial and boundary conditions in a ``\texttt{hard}" way could efficiently solve the incompressible NSEs by training the DNN without any labeled data~\cite{sun2020surrogate}. 
Although this ``\texttt{hard}" boundary condition enforcement is a novel approach which has the potential to be effective in solving various PDEs, it has not been fully explored.

In this work, we implement PINNs using the ``\texttt{hard}" boundary condition enforcement method as proposed in Ref.~\cite{sun2020surrogate}. 
We apply this approach to solve the incompressible, steady-state 2D NSEs for various systems with different geometries. In addition, we extend the analysis to a transient flow problem to demonstrate the broader applicability of the method. 
Specifically, we consider three representative  flow problems: (i) 2D lid-driven cavity flow (ii) flow past a circular obstacle in a pipe, and (iii) flow in a pipe with a sinusoidal boundary. 
The lid-driven case serves as a classical benchmark problem for validating any numerical methods. 
We investigate  steady flows for this case at three Reynolds numbers -low ($Re=100$), moderate ($Re=400$), and high ($Re=1000$) - to compare the performance of our PINN framework with conventional CFD results.
The second case, flow past a circular obstacle, exhibits a wide range of 
flow phenomena, such as boundary layer separation, wake formation etc. 
A PINN framework, which shows robust and accurate performance for this flow, 
can also be extended to more complex engineering and physical systems. We 
study this flow problem at $Re = $ 5, 20 and 40. 
The third case, on the other hand, has not been explored much in the context of PINNs. It plays a significant role in various natural and physiological processes, like conduit flow, stenotic and aneurysmal blood flow, urinary flow, etc. 
We first analyze the steady flow in a pipe with a sinusoidal boundary at different $Re$ values and subsequently extend the framework to study the transient flow in the same geometry with time-dependent inlet pressure. 
In addition to ``\texttt{hard}" boundary condition enforcement ~\cite{sun2020surrogate}, 
we incorporate a learning rate scheduler in our PINN framework. The scheduler algorithm dynamically adjusts the learning rate 
during the training whenever the loss function fails to decrease for a certain number of epochs. This strategy allows the network to take relatively larger steps in the early stages of training when the loss landscape is steep and progressively smaller steps as it approaches the minimum, leading to an improved convergence and stability. 
To the best of our knowledge, this is the first comprehensive study of hard-BC PINNs applied across a sequence of increasingly complex flow scenarios, culminating in a case relevant to real-world unsteady internal flows such as arterial or pulsatile transport.

The paper is organized as follows: we briefly discuss the methodology along with the basic principle of PINN, implementation of ``\texttt{hard}" boundary condition, learning rate scheduler algorithm and details of our CFD setup in Sec.\ref{sec:methodology}. Next, we present the results for the above-mentioned four problems in Sec.\ref{sec:result}. Finally, we summarize in Sec.\ref{sec:conclusion}. 

\section{Methodology}
\label{sec:methodology}


\subsection{Navier Stokes Equations}
\label{sec:nse}


The two-dimensional incompressible, laminar, steady-state Navier Stokes equations (NSEs) consist of mass conservation, $x^\prime$ and $y^\prime$ direction momentum conservation equations, as shown below, along with appropriate boundary 
conditions\footnote{Our case studies only involve the Dirichlet boundary conditions, and no Neumann boundary conditions will be considered.}.

\begin{eqnarray}
\frac {\partial u^\prime}{\partial x^\prime} +  \frac{\partial v^\prime}{\partial y^\prime} =  0 				  \label{eqn:mass_cons}\\
u^\prime \frac {\partial u^\prime}{\partial x^\prime} + v^\prime \frac {\partial u^\prime}{\partial y^\prime}  = 
- \frac{1}{\rho}	\frac {\partial p^\prime}{\partial x^\prime} + 
	\nu 	(\frac {\partial^2 u^\prime}{\partial x{^\prime}^2}  
		+ \frac {\partial^2 u^\prime}{\partial y{^\prime}^2}  )  \\
    u^\prime \frac {\partial v^\prime}{\partial x^\prime} + v^\prime \frac {\partial v^\prime}{\partial y}  = 
- \frac{1}{\rho}	\frac {\partial p^\prime}{\partial y^\prime} + 
	\nu 	(\frac {\partial^2 v^\prime}{\partial x{^\prime}^2}  
		+ \frac {\partial^2 v^\prime}{\partial y{^\prime}^2} )  
				  \label{eqn:momentum_cons}
\end{eqnarray}
where $u^\prime$ and $v^\prime$ are velocity components of the fluids in the $x^\prime$- and $y^\prime$- directions. $p^\prime, \rho$ and $\nu$ are the pressure, density and kinematic viscosity of the fluid, respectively. The governing equations \ref{eqn:mass_cons}-\ref{eqn:momentum_cons} can be non-dimensionalized using the following dimensionless variables:
\begin{equation}
x=\frac {x^\prime}{L}, ~ y=\frac {y^\prime}{L}, ~ u=\frac {u^\prime}{U^*}, ~ v=\frac {v^\prime}{U^*}, ~ p =\frac {p^\prime}{\rho{U^*}^2}, ~ Re= \frac {{U^*}L}{\nu}
\label{eqn:dim_scale}
\end{equation}
where $L$ and $U^*$ denote characteristic length and characteristic velocity, respectively. $Re$ is the Reynolds Number of the system, a dimensionless quantity, which determines the turbulent versus laminar condition in a flow. At high Reynolds numbers, the inertial force dominates over the viscous forces, leading to a turbulent state of flow. Using the above non-dimensional scales (Equation~\ref{eqn:dim_scale}) into the Equations~\ref{eqn:mass_cons}-\ref{eqn:momentum_cons}, we obtain the non-dimensional form of NSEs as follows:
\begin{eqnarray}
\frac {\partial u}{\partial x} +  \frac{\partial v}{\partial y} =  0 
				  \label{eqn:nse_non_dim1}			  \\
u \frac {\partial u}{\partial x} + v \frac {\partial u}{\partial y}  = 
- 	\frac {\partial p}{\partial x} + 
	\frac {1}{Re} 	(\frac {\partial^2 u}{\partial x^2}  
		+ \frac {\partial^2 u}{\partial y^2}  )  
		\label{eqn:nse_non_dim2}	\\
    u \frac {\partial v}{\partial x} + v \frac {\partial v}{\partial y}  = 
- \frac {\partial p}{\partial y} + 
	\frac{1}{Re} 	(\frac {\partial^2 v}{\partial x^2}  
		+ \frac {\partial^2 v}{\partial y^2} )  
				  \label{eqn:nse_non_dim3}
				  \end{eqnarray}
				  
\subsection{Neural Network}
\label{sec:nn}

A neural network is a machine learning algorithm inspired by the functioning of neurons in the human brain. It consists of multiple layers of interconnected units called neurons. These layers include an input layer that receives the function's input, one or more hidden layers where most of the computation is done, and an output layer that produces the output of the function.
Every neuron in a layer ``$l$" is connected to the neurons in the next layer. The connections are associated with weights ($w_{ji}^{(l)}$)  and each neuron has a bias ($b_{j}^{(l)}$). The output of a single $j^{th}$ neuron ($a^{(l)}_{j}$) in the $l^{th}$ layer depends on the input it receives from the previous layer ($a_{i}^{(l-1)}$), given by the equation:

\begin{equation}
  a_{j}^{(l)} = \sigma\left(\sum_{i} w_{ji}^{(l)}a_{i}^{(l-1)} + b_{j}^{(l)}\right),
\end{equation}
where $\sigma$ is called the activation function, which introduces non-linearity in the NN via nonlinear functions, e.g., sigmoid, tanh, ReLU, Softmax, etc. For a network with L layers, the output of the network ($\boldsymbol{f}_{NN}(\boldsymbol{x})$) is given by:
\begin{equation}
 \left(\boldsymbol{f}_{NN}(\boldsymbol{x})\right)_j = a^{(L)}_{j} = \sigma\left(\sum_{i} w_{ji}^{(L)}a_{i}^{(L-1)} + b_{j}^{(L)}\right),
\end{equation}
where $a_{i}^{(L-1)}$ is the output of the $i^{th}$ neuron in the $(L-1)^{th}$ layer, which can be computed recursively using the output of the previous layer.


Neural Networks have the capability to approximate any function, given sufficient neurons and layers \cite{Cybenko1989ApproximationBS}. The network learns the function through adjustment of the weights and biases. The weights and biases are initialized randomly and then updated during the training process such that a certain loss function is minimized. The loss function is a measure of the error in the prediction of the network. The gradients are calculated using an algorithm called the Backpropagation algorithm \cite{rumelhart1986learning}, which applies the chain rule.
With this gradient information, the weights and biases are updated using the gradient descent algorithm as follows:
\begin{equation}
    w_{ji}^{(l)} \rightarrow w_{ji}^{(l)} - \eta \frac{\partial \mathcal{L}}{\partial w_{ji}^{(l)}};  ~~~ b_{j}^{(l)} \rightarrow b_{j}^{(l)} - \eta \frac{\partial \mathcal{L}}{\partial b_{j}^{(l)}},
\end{equation}
where $\eta$ is the learning rate (LR), and $\mathcal{L}$ is the loss function. There are several algorithms used by the optimizers for updating the weights and biases at every iteration. Most commonly used optimizers include the Stochastic Gradient Descent (SGD), Adaptive Moment Estimation (Adam), Limited-memory Broyden-Fletcher-Goldfarb-Shanno Algorithm (LBFGS), Root Mean Square Propagation (RMSprop), etc. 
While training a neural network, the input features of a labeled dataset are fed to the network, and the network learns the function that maps the input features to the output labels. This is done by first calculating $\boldsymbol{f}_{NN}(\boldsymbol{x})$ for the dataset and then evaluating the differentiable loss function $\mathcal{L}$. The loss function is then minimized by computing its gradient with respect to the weights and biases and updating them using the gradient descent algorithm. This process is repeated for several epochs until the loss function converges to a minimum value.
The learning rate $\eta$ is a hyperparameter that controls the step size at each iteration while approaching a minimum of the loss function. A small $\eta$ may lead to slow convergence, while an excessively large learning rate may cause the
divergence or oscillation, particularly as the loss landscape flattens and gradients 
become vanishing during training.
Therefore, selecting an appropriate learning rate is critical for stable and efficient training of the PINN.

\subsection{Physics-Informed Neural Networks}

\begin{figure}[!htb]
    \centering
    \includegraphics[width=0.9\textwidth]{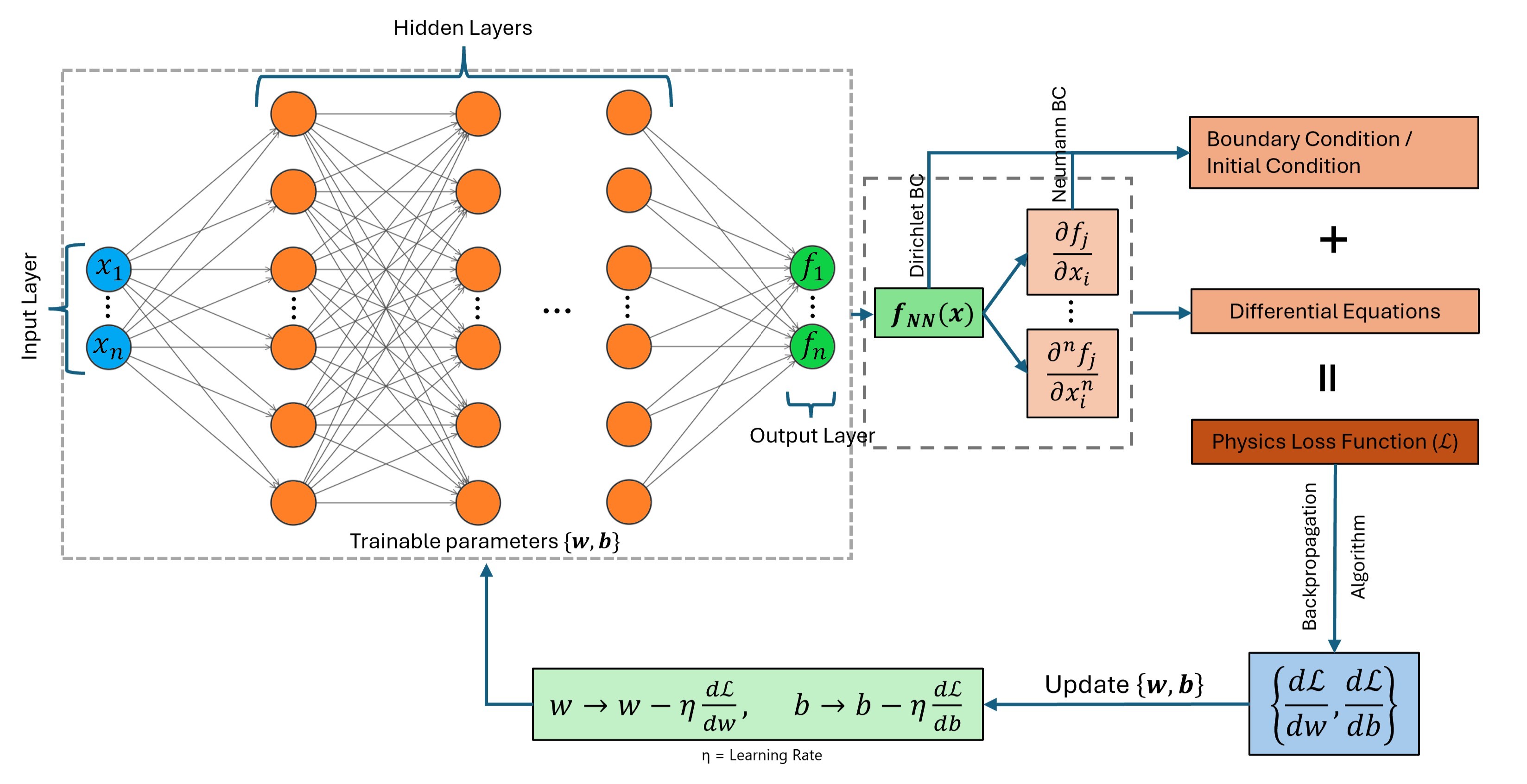}
    \caption{The architecture of a Physics-Informed Neural Network (PINN).}
    \label{fig:pinn}
\end{figure}

PINN uses physics-based loss functions to train the neural network. This is done by incorporating the governing equations of the system (residuals of the PDEs) as a term in the loss function. Let a PDE be given as $\mathcal{D}(f(x), x) = 0$, where $f(x)$ is the solution to the PDE. Then, the square of $\mathcal{D}$ can be added to the Loss function of the neural network since for any function that satisfies the PDE, the value of $\mathcal{D}(f(x), x)$ will be zero.
The Mean Square Error (MSE) of the value of the function at the boundary can also be added to the loss function as the loss due to the Dirichlet Boundary conditions. In a similar way, for the Neumann Boundary conditions, one can calculate the derivative of the function at the boundary and add its MSE to the loss function. The total loss function is then given by:
\begin{equation}
    \mathcal{L}_{PINN} = \mathcal{L}_{data} + \mathcal{L}_{PDE} + \mathcal{L}_{\text{Dirichlet}} + \mathcal{L}_{Neumann},
    \label{eq:loss}
\end{equation}
where, $\mathcal{L}_{data}$ is the loss term due to the given labeled data, $\mathcal{L}_{PDE}$ is the loss function due to the governing PDEs, $\mathcal{L}_{\text{Dirichlet}}$ is the loss function due to the Dirichlet Boundary conditions, and $\mathcal{L}_{Neumann}$ is the loss function due to the Neumann Boundary conditions\footnote{For unsteady-state problems, the initial conditions also contribute to the loss function denoted as $\mathcal{L}_{IC}$}. Thus, the network is now trained to use this loss function. All the derivatives required are calculated using the automatic differentiation capabilities of the deep learning frameworks. To give more importance to a particular loss term, different weights can be assigned to each term. The governing equations for a 2D laminar incompressible fluid in a stationary state are given by the non-dimensional NSEs as mentioned in Equations~\ref{eqn:nse_non_dim1}-\ref{eqn:nse_non_dim3}. 
Therefore, the function to approximate is given by: 
$\boldsymbol{f}_{NN}(x, y) = (u , v, p)$ where $u, v$ are the x and y component velocity field, $p$ is the pressure field for a particular  Reynolds number 
$\text{Re}$. Using Equations~\ref{eqn:nse_non_dim1}-\ref{eqn:nse_non_dim3}, under suitable boundary conditions, one can calculate the loss function for the PINN can be calculated and utilized to train the network to approximate the solution.
The schematic of our PINN architecture is shown in Figure \ref{fig:pinn}.

\subsection{Hard Boundary Condition}

For a vanilla data-free PINN, the loss function is estimated from several contributions as shown in Equation~\ref{eq:loss}, apart from $\mathcal{L}_{data}$. In the traditional CFD cases, the solution is built upon the boundary conditions. In contrast, PINN is trained to approximate the solution by optimizing the loss function constructed to enforce both the governing PDEs and the boundary conditions. Nevertheless, the network may not satisfy the boundary conditions precisely in some cases.
NSEs, being the composite of three differential equations, may include several boundary conditions like no-slip boundary conditions along the walls or slip conditions with specified velocity or pressure at the inlet and outlet, depending on the system.
Thus, it becomes challenging for the network to simultaneously minimize all the losses for different conditions.
To overcome this, L. Sun et al. \cite{sun2020surrogate} proposed the imposition of boundary conditions in a ``\texttt{hard}" manner such that the output function itself satisfies the BCs. 
 This is achieved by defining the output function $f(x)$ as
\begin{equation}
    \label{eq:hard_bc}
    f(x) = f_{\text{NN}}(x) A(x) + B(x),
\end{equation}
where $f_{\text{NN}}(x)$ is the output of the neural network, and $A(x)$ \& $B(x)$ are chosen in such a way that the boundary condition for $f(x)$ is automatically satisfied. By doing so, one can eliminate the loss corresponding to the Dirichlet Boundary condition in the Loss function given in Equation \ref{eq:loss} and also ensure that the boundary conditions are exactly satisfied. 
The choice of $A(x)$ and $B(x)$ depends on the type of boundary conditions. $A(x)$ can be chosen as any function that is zero at the boundary and non-zero elsewhere, while $B(x)$ can be chosen as the value of the velocity at the boundary. This way, the output function $f(x)$ will have the desired value at the boundary, while in the bulk of the domain, the output will be determined by the neural network. While choosing $A(x)$ and $B(x)$, to ensure fast convergence of the network, one must consider simple functions that can be easily computed and do not introduce any additional complexity to the problem. For more complex boundary conditions, one can train a separate neural network to learn the functions $A(x)$ and $B(x)$, which can then be used to construct the output function $f(x)$.

In this work, we have adapted this method to solve the 2D Navier-Stokes equations for different systems. We are using three sub-neural networks to approximate the $u, v, p$ fields. For each network, the output of the neural network will be multiplied by a suitable function $A(x)$ and added to another function $B(x)$ to satisfy the boundary conditions. The loss function given in Equation \ref{eq:loss} becomes:
\begin{equation}
   \mathcal{L}_{Data-free-PINN}^{Hard-BC} = \mathcal{L_{\text{PDE}}} = 
   \Lambda^{(u)}_{\text(mom)}\mathcal{L}_{\text{mom}}^{(u)} + \Lambda^{(v)}_{\text(mom)}\mathcal{L}_{\text{mom}}^{(v)} + \Lambda^{(c)}_{\text{cont}}\mathcal{L_{\text{cont}}}
\end{equation}
where $\mathcal{L}_{\text{mom}}^{(u)}$, $\mathcal{L}_{\text{mom}}^{(v)}$ and $\mathcal{L_{\text{cont}}}$ are the loss terms corresponding to the residual 
of the Navier-Stokes equations~\ref{eqn:nse_non_dim1}-\ref{eqn:nse_non_dim3}, and $\Lambda$'s are hyperparameters which can be tuned to give more importance to a particular loss term.
Let $\mathcal{X} = \{(x, y)\}$ be the set of collocation points selected in the domain; then the loss terms are defined as:
\begin{eqnarray}
    \mathcal{L}_{\text{mom}}^{(u)} = \frac{1}{|\mathcal{X}|} \sum_{i \in \mathcal{X}} \left[ u \frac{\partial u}{\partial x} + v \frac{\partial u}{\partial y} + \frac{\partial p}{\partial x} - \frac{1}{\text{Re}} \left( \frac{\partial^2 u}{\partial x^2} + \frac{\partial^2 u}{\partial y^2} \right) \right]^2_{i} \nonumber \\
    \mathcal{L}_{\text{mom}}^{(v)} = \frac{1}{|\mathcal{X}|} \sum_{i \in \mathcal{X}} \left[ u \frac{\partial v}{\partial x} + v \frac{\partial v}{\partial y} + \frac{\partial p}{\partial y} - \frac{1}{\text{Re}} \left( \frac{\partial^2 v}{\partial x^2} + \frac{\partial^2 v}{\partial y^2} \right) \right]^2_{i} \\
    \mathcal{L}_{\text{cont}} = \frac{1}{|\mathcal{X}|} \sum_{i \in \mathcal{X}} \left[ \frac{\partial u}{\partial x} + \frac{\partial v}{\partial y} \right]^2_{i} \nonumber
   				  \label{eqn:loss_hard}
\end{eqnarray}

\subsection{Learning rate scheduler}
\label{sec:scheduler}
In this analysis, we train the model using the Adam optimizer \cite{kingma2017adammethodstochasticoptimization}, whose effectiveness depends 
on the choice of learning rate. As discussed in Section~\ref{sec:nn}, a very small or large $\eta$ may lead to slow convergence or induce divergence. 
To optimize the convergence and stability of the training process, we adopt a  learning rate scheduler algorithm in our PINN framework.
A learning rate scheduler dynamically adjusts the learning rate ($\eta$) during training. For example, if the loss does not decrease for a certain number of epochs, the scheduler reduces the learning rate by a factor $\beta$ (where $0<\beta<1$). This helps the network take relatively larger steps when the optimization landscape is steep and progressively smaller steps as it approaches the minimum, leading to improved convergence and stability.
The scheduler updates the learning rate after $n_p$ (\texttt{patience}) iterations, if no improvement in the loss is observed. The learning rate updates as:
\begin{equation}
\eta_{i+1} = max \big( \eta_i . \beta, \eta_{min}\big)
\end{equation}
where $\eta_i$ is the learning rate at the current epoch, 
$\beta$ is the learning rate reduction factor ($0<\beta<1$) and $\eta_{\min} $ is the minimum allowable learning rate. 
and set to $10^{-6}$ or $10^{-8}$ depending on the flow problem. In this work, we start the training with a learning rate $10^{-2}$ for steady cases and $10^{-3}$ for transient flow. It is worth mentioning that using a large learning rate at the initial stage of training can sometimes help the model to explore the loss landscape more effectively.

\subsection{Network hyperparameters}
\label{sec:hyper}

The accuracy of the PINN framework critically depends on the choices of the network hyperparameters. We use 3 different  neural sub-networks to predict the horizontal velocity $u$, vertical velocity $v$, and pressure $p$ fields separately. For each of these networks, we keep the number of hidden layers and neurons per layer the same. The number of hidden layers and neurons per layer for the 4 different flow problems are mentioned in Table~\ref{tab:hyperparameters_revised}. We use the Swish\cite{ramachandran2017searchingactivationfunctions} activation function \texttt{(SiLU)} for all the cases. The weights are initialized using the Xavier\cite{he2015delvingdeeprectifierssurpassing} initialization method.

The batch size is the number of collocation points considered in the domain. The collocation points are considered in different ways for different geometries. More details on it are discussed in the later sections. We train the models for $3 \times 10^5$ iterations for all the cases. Although in most cases the model converges much earlier, we consider a higher number of iterations to ensure the convergence. To update the weights and biases, the choice of optimizer is crucial. Several studies  \cite{raissi2019deep,RAO2020207,mca28050102} have reported that using Adam optimizer at initial stages of training and then switching to the LBFGS optimizer at later stages can lead to better convergence. However, the LBFGS optimizer is not suitable for large batch sizes due to its high memory requirement, and computationally expensive nature. We employ the Adam optimizer~\cite{kingma2017adammethodstochasticoptimization} in our PINN framework while also employing a learning rate scheduler as discussed in Section~\ref{sec:scheduler}.

\begin{table}[!htb]
\centering
\begin{tabular}{@{}lcccc@{}}
\toprule
\textbf{Hyperparameter} & \textbf{Lid-Driven Cavity} & \textbf{Circular Obstacle} & \textbf{Sine Pipe} & \textbf{Sine Pipe } \\
\textbf{} & (Steady-state) & (Steady-state) &(Steady-state) & (Transient) \\ \midrule
Batch Size              & $10000$            & $2 \times 10000$  & $1000$             & $10000$                        \\
Num. Iterations         & $3 \times 10^5$            & $3 \times 10^5$            & $3 \times 10^5$    & $3 \times 10^5$                \\
Initial LR              & $10^{-2}$                  & $10^{-2}$                  & $10^{-2}$          & $10^{-3}$                      \\ \midrule
\textit{\textbf{Network Architecture:}}                                                                                \\
No. of Hidden Layers    & 3                          & 4                          & 3                  & 4                              \\
Neurons per Layer       & 128                        & 100                        & 64                 & 200                            \\ \midrule
\textit{\textbf{Loss Weightage:}}                                                                                     \\
u-momentum loss         & 1                          & 1                          & 1                  & 1                              \\
v-momentum loss         & 1                          & 1                          & 1                  & 1                              \\
\multirow{3}{*}{Continuity loss} & 1 (Re=100)        & \multirow{3}{*}{10}        & \multirow{3}{*}{1} & \multirow{3}{*}{10}            \\
                        & 10 (Re=400)                &                            &                    &                                \\
                        & 50 (Re=1000)               &                            &                    &                                \\ \midrule
\textit{\textbf{Learning rate scheduler:}}                                                                                       \\
Patience                & 100                        & 100                        & 100                & 100                            \\
Factor                  & 0.95                       & 0.8                        & 0.95               & 0.95                           \\
Min. LR                 & $10^{-8}$                  & $10^{-6}$                  & $10^{-8}$          & $10^{-8}$                      \\ \bottomrule
\end{tabular}
\caption{Hyperparameter settings for the neural network models. The Batch size is the number of collocation points chosen in the domain. For most cases we have chosen fixed collocation points, except for the Sine Pipe steady-state case, where we have randomly chosen collocation points at every iteration.}
\label{tab:hyperparameters_revised}
\end{table}

For more complex problems it has been observed that the model converges better if the continuity loss is given more weightage. Additionally, for the case of lid-Driven Cavity flow and transient flow in a sinusoidal boundary pipe, while training in GPU it has been observed that the model converges better if we disable NVIDIA's cuDNN library. This can be attributed to the fact that cuDNN uses non-deterministic algorithms to optimize the performance. Therefore disabling it makes gradient computation more stable and deterministic leading to lower final loss values. Particularly for the lid-driven cavity flow at $Re=1000$, we use float64 precision for training to get better accuracy. However, float64 precision increases the memory requirement and training time significantly. The hyperparameter settings for all the four flow problems are summarized in Table~\ref{tab:hyperparameters_revised}.

\subsection{Outline of CFD computations}

\begin{figure}[!htb]
    \centering
                \includegraphics[width=0.4\textwidth]{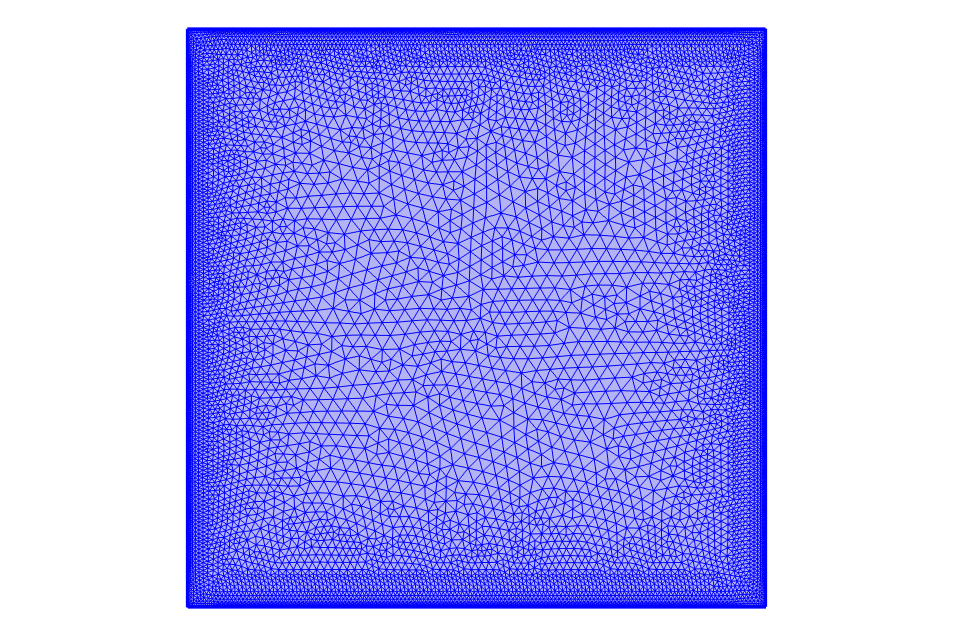}
                \vspace{+0.5 cm}
                    \includegraphics[width=0.4\textwidth]{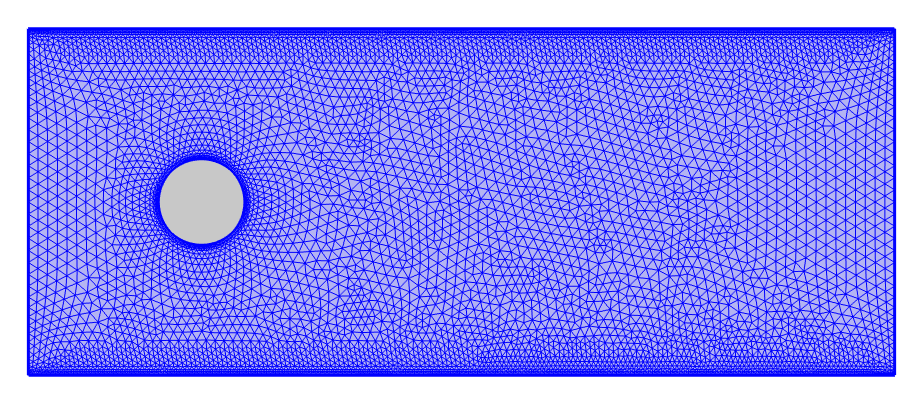} 
                    \includegraphics[width=0.8\textwidth]{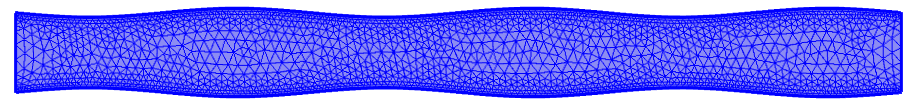}

    \caption{Schematic of mesh generation in CFD for the three benchmark cases - the 2D lid-driven cavity flow (top left),  flow passing a circular obstacle in a pipe (top right) and flow in a pipe with sinusoidal boundary (bottom).
}
                  \label{fig:mesh}
\end{figure}

To validate the predicted results by our PINN framework simulations, we also conduct the CFD studies using COMSOL Multiphysics 5.3, which is based on Galerkin Finite Element Method. Within COMSOL, we implement the non-dimensional, steady-state NSEs as presented in equations~\ref{eqn:dim_scale}-\ref{eqn:nse_non_dim1} under the 
appropriate boundary conditions. It may be noted that COMSOL employs the elimination-based Parallel Direct Sparse Solver (PARDISO) algorithm to solve the system of linear equations arising from the linearization of the nonlinear NSEs. For the CFD simulations, we use physics-controlled mesh, which essentially discretizes the domain into rectangular and triangular elements, as shown in Fig.\ref{fig:mesh}, with required mesh refinement depending on applied boundary conditions. For example, elements are finer and rectangular near the boundary where no-slip boundary conditions are applied. However, the inlet/outlets and the main domain are meshed with triangular elements.

\begin{table}[!htb]
\begin{center}
\begin{tabular}{||c||c||c||c||c||c||c||c||c||}
\hline\hline
Grids & Extremely  & Coarse & Normal & Fine & Finer & Extra & Extremely \\
	& Coarse &  & 	&  &  & Fine & Fine \\
\hline \hline
Number  &  &  &  &  &  &  &  \\
 of Mesh & 222 & 1094 & 1610 & 2646 & 6884 & 17666 & 26910 \\
 Elements &  &  &  &  &  &  &  \\
\hline\hline
$|u_{\text{avg}}|$ at &  &  &  &  &  &  &  \\
 horizontal & 0.1170 & 0.1266 & 0.1283 & 0.1289 & 0.1287 & 0.1288 & 0.1289 \\
 centerline &  &  &  &  &  &  &  \\
\hline \hline
\end{tabular}
\caption{Grid independent study  for lid-driven cavity flow at $Re = 100$ in CFD.}
\label{tab:cfd_grid}
\end{center}
\end{table}

\begin{table}[!htb]
\begin{center}
\begin{tabular}{||c||c||c||c||c||}
\hline\hline
\textbf{CFD Parameters} & \textbf{Lid-driven Cavity} & \textbf{ circular obstacle} & \textbf{Sine Pipe }& \textbf{Sine Pipe }\\ 
 &\textbf{(Steady-state)}  & \textbf{(Steady-state)}  & \textbf{(Steady-state)}& \textbf{(Transient)}\\ 
\hline\hline
 Domain elements		& 	26310 & 55736		 & 	149448	& 150779 \\
 	\hline
 Boundary elements		& 600	 & 	1044	 & 		2632 & 2889 \\
\hline
 Total elements			& 26910	 & 		56780 & 	152080	& 153668 \\
\hline
 Minimum element size	& 1.5$\times 10^{-4}$	 & 	 2.0$\times 10^{-5}$	 & 		 2.02$\times 10^{-5}$ &  2.29$\times 10^{-5}$\\
\hline
 Maximum element size	& 0.013	 & 	0.01	 & 	0.0101	& 0.0103 \\
\hline
 Maximum element 		& 	1.08 & 	1.1	 & 1.1	& 1.1	  \\
growth rate 	 		& 	 & 		 & 		&  \\
\hline
 Curvature factor		& 0.25	 & 	0.2	 & 	0.2	& 0.21 \\
\hline
 Resolution of 			& 	1 & 1		 & 	1	& 1 \\
narrow regions 			& 	 & 		 & 		&   \\
\hline
 Fully coupled & \multicolumn{4}{c||}{Newton (automatic)}	  \\
nonlinear method	& \multicolumn{4}{c||}{ }	  \\
\hline
Solver      & 	\multicolumn{4}{c||}{Parallel direct sparse
solver interface   }	  \\
      & 	\multicolumn{4}{c||}{ (PARDISO) }	  \\
\hline\hline
\end{tabular}
\caption{Parameter values considered in CFD simulations for our chosen scenarios.}
\label{tab:cfd_param}
\end{center}
\end{table} 

We also perform the grid convergence study to check the credibility of the CFD solution.
To illustrate the grid independence of the current analysis, we present the average velocity along the horizontal centerline for the lid-driven cavity flow at different mesh resolutions in Table~\ref{tab:cfd_grid}. For our work, we consider the extremely fine grids for CFD simulations. Additionally, the key COMSOL simulation parameters are summarized in Table~\ref{tab:cfd_param}.

Further, we study the accuracy of our PINN 
 model by evaluating the $L_2$ normalized error ($\epsilon_{L_2}$) of velocity components and pressure, which is defined as: 
 \begin{equation}
\epsilon_{L_2} = \sqrt{\frac{\sum_i(u_{i}^{CFD} - u_{i}^{NN})^2}{\sum_i(u_i^{CFD})^2}}
    \label{eq:l2}
\end{equation}		
where $u_i$ denotes the $x, y$ velocity component and pressure, and i denotes the points in the mesh generated in CFD.

\section{Results}
\label{sec:result}

In this section, we discuss the PINN results obtained for the four benchmark scenarios along with the implementation of boundary conditions, training methodology and detailed comparison with the corresponding CFD simulations.

\subsection{Steady flow in 2D lid-driven square cavity}

\begin{figure}[!htb]
    \centering
    \includegraphics[width=0.6\textwidth]{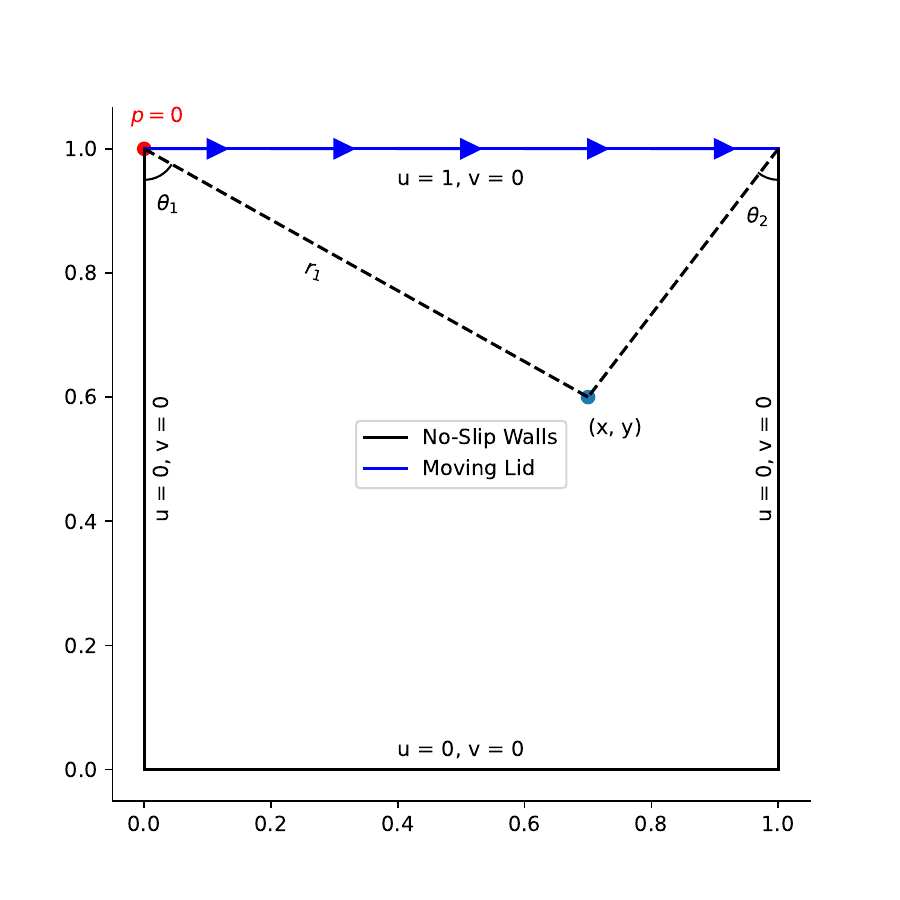}
    \caption{The geometry of the 2D square cavity with the top wall moving at a constant horizontal velocity ($u=1, v=0$). The bottom and side walls are considered to be no-slip walls. }
    \label{fig:lid_cavity_geo}
\end{figure}

The two-dimensional (2D) lid-driven cavity flow is a well-established benchmark problem for evaluating the accuracy and robustness of any new numerical techniques. We consider a 2D square cavity with a dimensionless side length of 
unity and impose a boundary condition of horizontal sliding ($u=1$) at the top wall. The flow is thus characterized by the non-dimensional parameter Reynolds number, which is defined in Sec.~\ref{sec:nse}. The geometry of the system and the boundary conditions are shown in Figure \ref{fig:lid_cavity_geo}.
Apart from the top wall, all other walls are considered to be under no-slip 
condition with zero velocity. We have implemented the same boundary conditions along with the pressure point constraint in our CFD simulations as well.

\begin{figure}[!htb]
    \centering
        \includegraphics[width=0.8\textwidth]{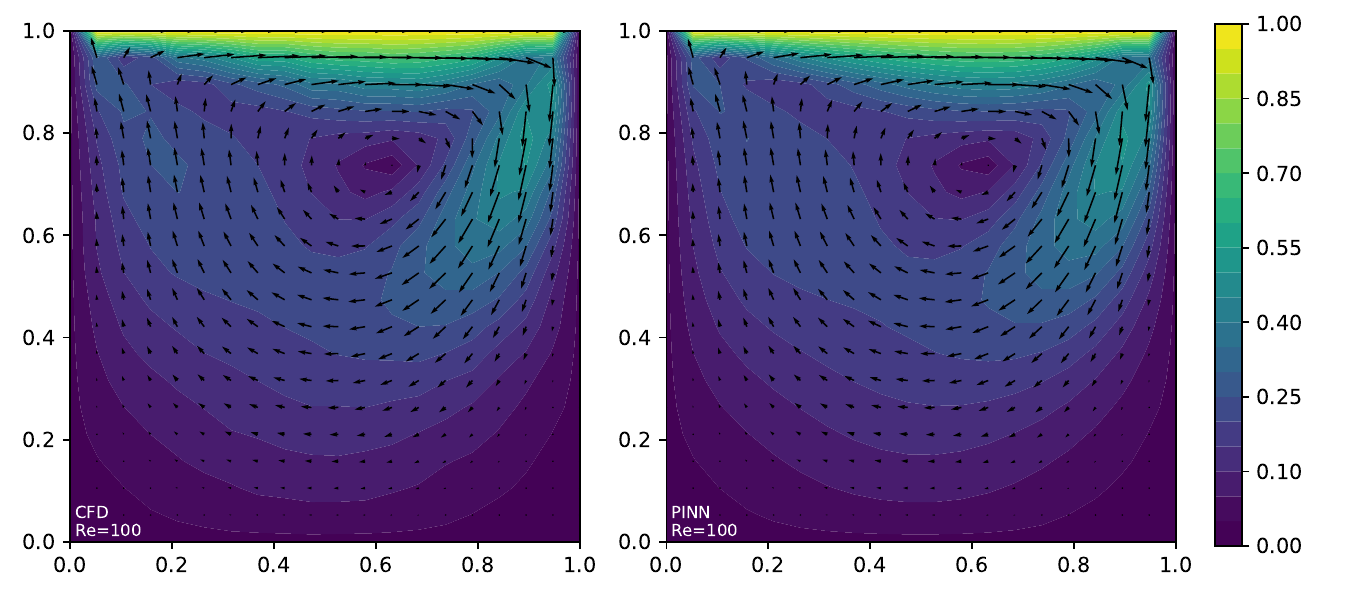}
    \caption{Velocity field for the 2D lid-driven cavity , with $Re = 100$,  obtained from CFD (left) and PINN (right). 
    Arrows represent velocity vectors, and color indicates velocity magnitude.
    }
        \label{fig:lid_driven_vel_field_100}  
\end{figure}

As discussed in Section \ref{sec:hyper}, we employ three sub-networks to model the horizontal velocity, vertical velocity, and pressure fields. Each sub-network consists of 
three hidden layers, with 128 neurons per layer.
The network architecture and optimized hyperparameters are enlisted in Table~\ref{tab:hyperparameters_revised} in Sec.~\ref{sec:hyper}.  
The hard boundary conditions (Hard-BCs) are imposed by taking the $x$, $y$ components of velocity and pressure as: 
\begin{eqnarray}
 u(x, y) = u_{\text{NN}}(x, y) x(1 - x) y (1 - y) +  \frac{4}{\pi^2} \theta_1 \theta_2 y \label{eqn:u1}\\
v(x, y) = v_{\text{NN}}(x, y) x(1 - x)y (1 - y) \\
p(x, y) = p_{\text{NN}}(x, y) r_1 \label{eqn:p}
\end{eqnarray}
where $\theta_1 = \tan^{-1}\left(\frac{x}{1- y}\right)$, $\theta_2 = \tan^{-1}\left(\frac{1 - x}{1 - y}\right)$ and $r_1 = \sqrt{x^2 + (1-y)^2}$. 
It is evident that these equations (Eq.\ref{eqn:u1}-\ref{eqn:p}) satisfy the boundary conditions, as shown in Figure \ref{fig:lid_cavity_geo}, for the 2D 
lid-driven cavity. More details about the choices and implementation of these appropriate Hard-BC functions are given in Appendix~C. The collocation points are chosen to be $100 \times 100$ uniformly spaced grid points in the computational domain.
The points around the boundary are avoided because of the discontinuous horizontal velocity at the two upper corners. This is done by selecting points in the domain $0.01 \leq x \leq 0.99$ and $0.01 \leq y \leq 0.99$. 
This does not affect the solution close to the boundary as the Hard-BCs automatically ensure that the solution at the boundary is exactly equal to the boundary conditions. 
For illustration purposes, we present the results for three Reynolds numbers: $Re = 100$, 400 and 1000.

\begin{figure}[!htb]
    \centering
        \includegraphics[width=0.8\textwidth]{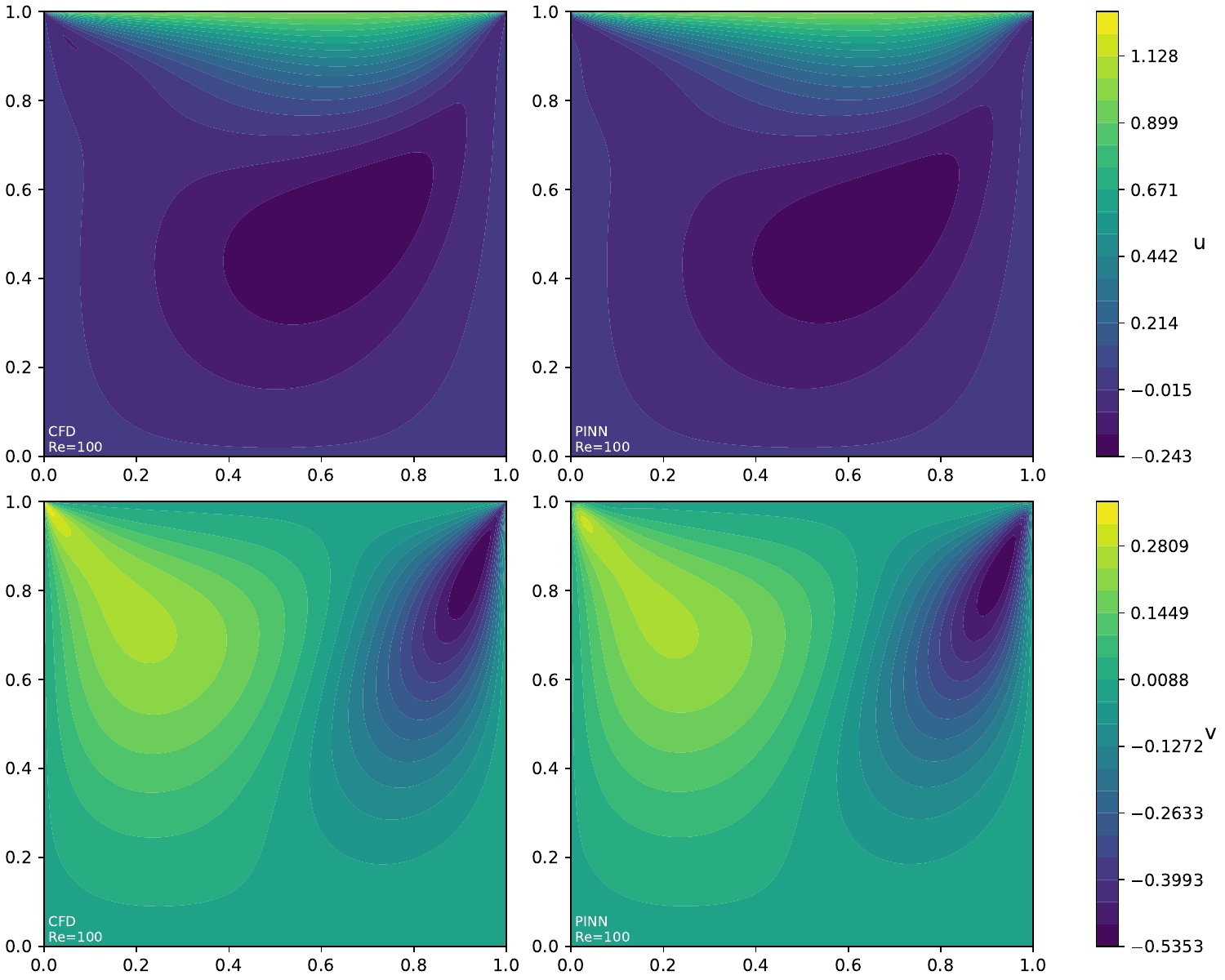}
    \caption{Comparison of the results obtained from PINN (right) and CFD simulations (left) for the lid-driven cavity with Re = 100. 
    The top and bottom panels  correspond to the $x$- and $y$-components of velocity ($u, v$).
    }
    \label{fig:lid_driven_uvp}
\end{figure}

In Figure \ref{fig:lid_driven_vel_field_100}, 
we compare the velocity field obtained from CFD simulation (left) and our PINN framework (right) for low Reynolds Number $Re = 100$, where the viscous force dominates over the inertial force.
From the velocity profiles, we observe that 
the velocity approaches 1 and 0 near the top wall and the remaining no-slip walls, respectively. The vortex position and shape are also well captured by the PINN model. The comparisons of the horizontal velocity field (u) and vertical velocity field (v) for this model are shown in the top and bottom panels of Figure \ref{fig:lid_driven_uvp}. These figures also indicate that the results obtained from the PINN frameworks (right column) are in good agreement with those from the CFD simulations (left column).

\begin{figure}[!htb]
    \centering
    \begin{subfigure}[b]{0.49\textwidth}
        \centering
        \includegraphics[width=\textwidth]{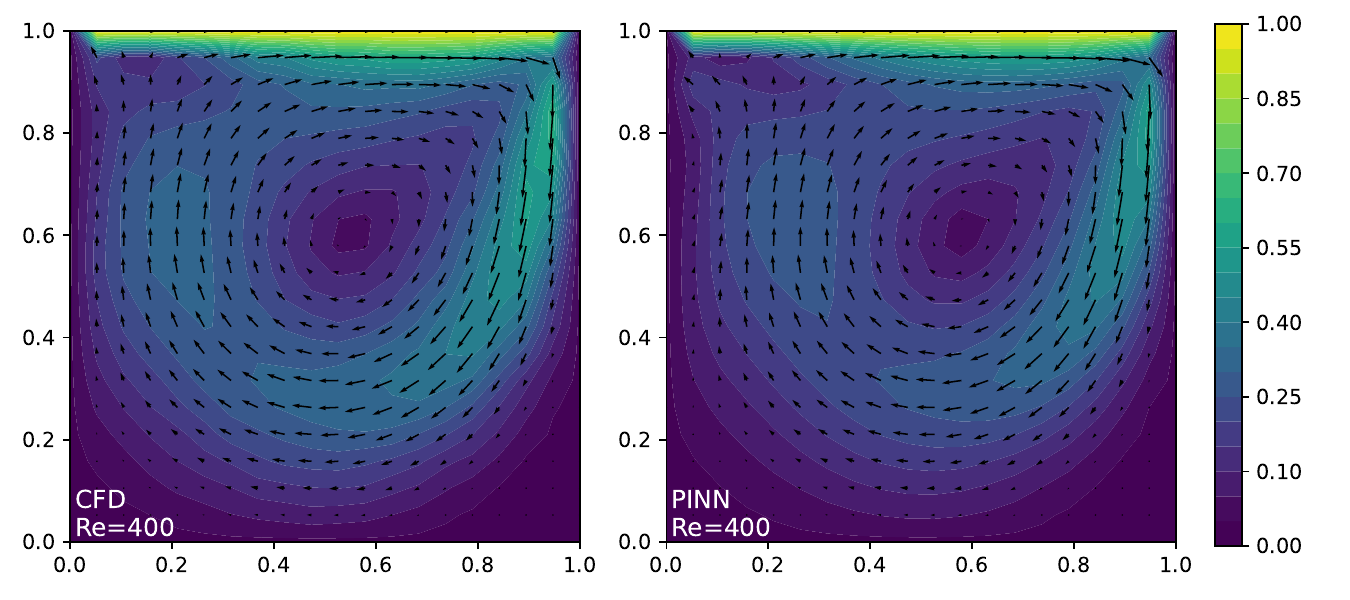}
    \end{subfigure}
    \begin{subfigure}[b]{0.49\textwidth}
        \centering
        \includegraphics[width=\textwidth]{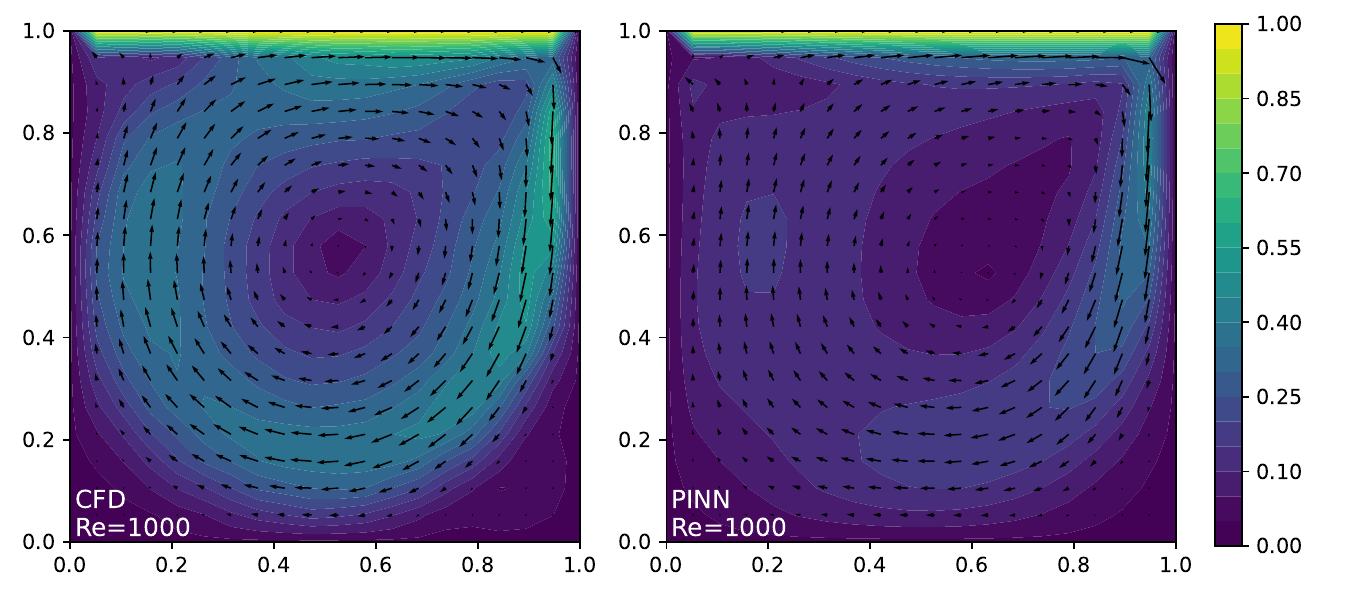}
    \end{subfigure}
    \caption{Velocity field for the 2D lid-driven cavity obtained from CFD and PINN with Re = 400 (first two columns) and 1000 (last two columns). Arrows represent velocity vectors, and color indicates velocity magnitude.}
    \label{fig:lid_driven_vel_field_400_1000}
\end{figure}

\begin{figure}[!htb]
    \centering
    \begin{subfigure}[b]{0.48\textwidth}
        \centering
        \includegraphics[width=\textwidth]{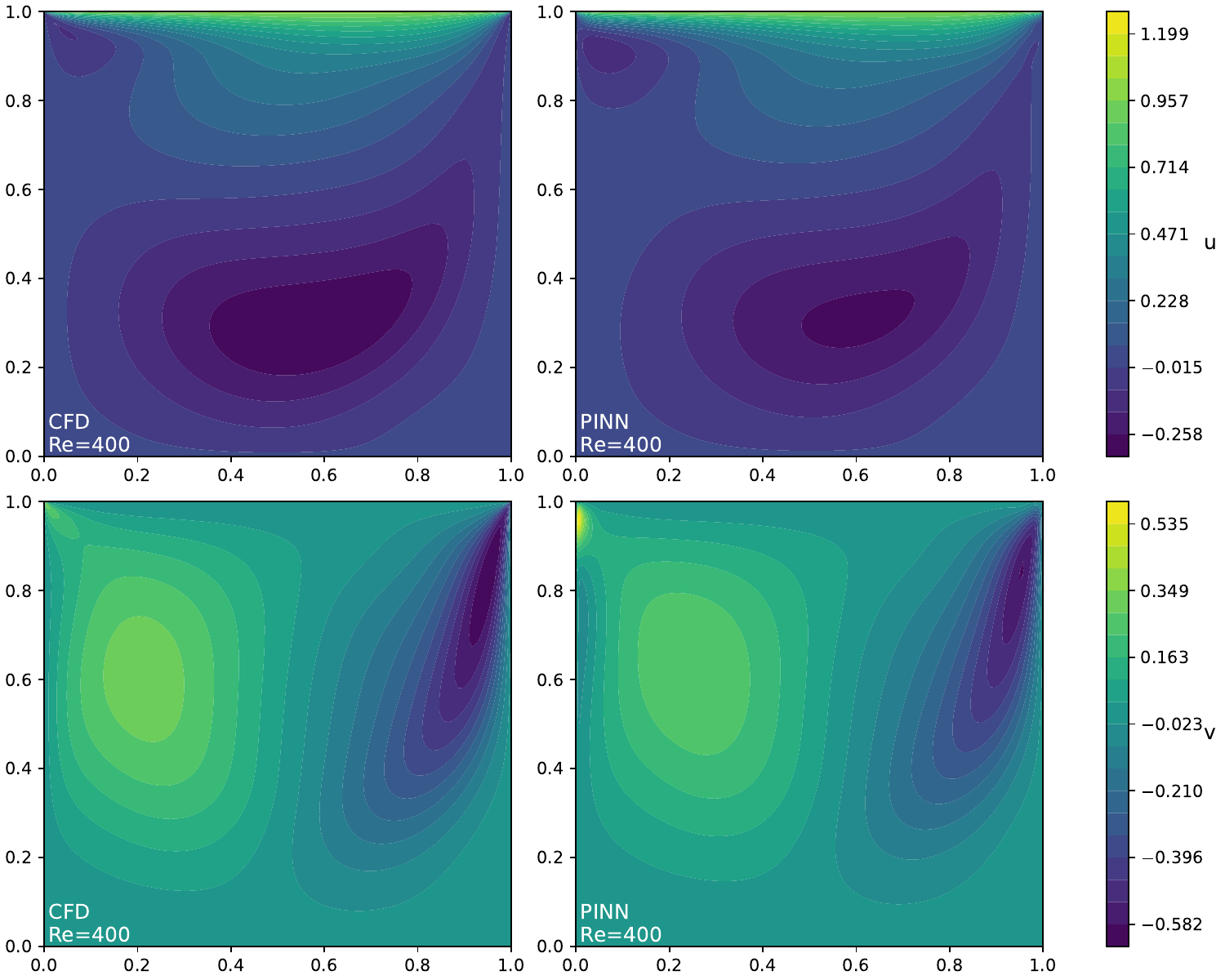}
        \label{fig:lid_driven_uvp_400}
                \caption{Re=400}
    \end{subfigure}    
    \begin{subfigure}[b]{0.48\textwidth}
        \centering
        \includegraphics[width=\textwidth]{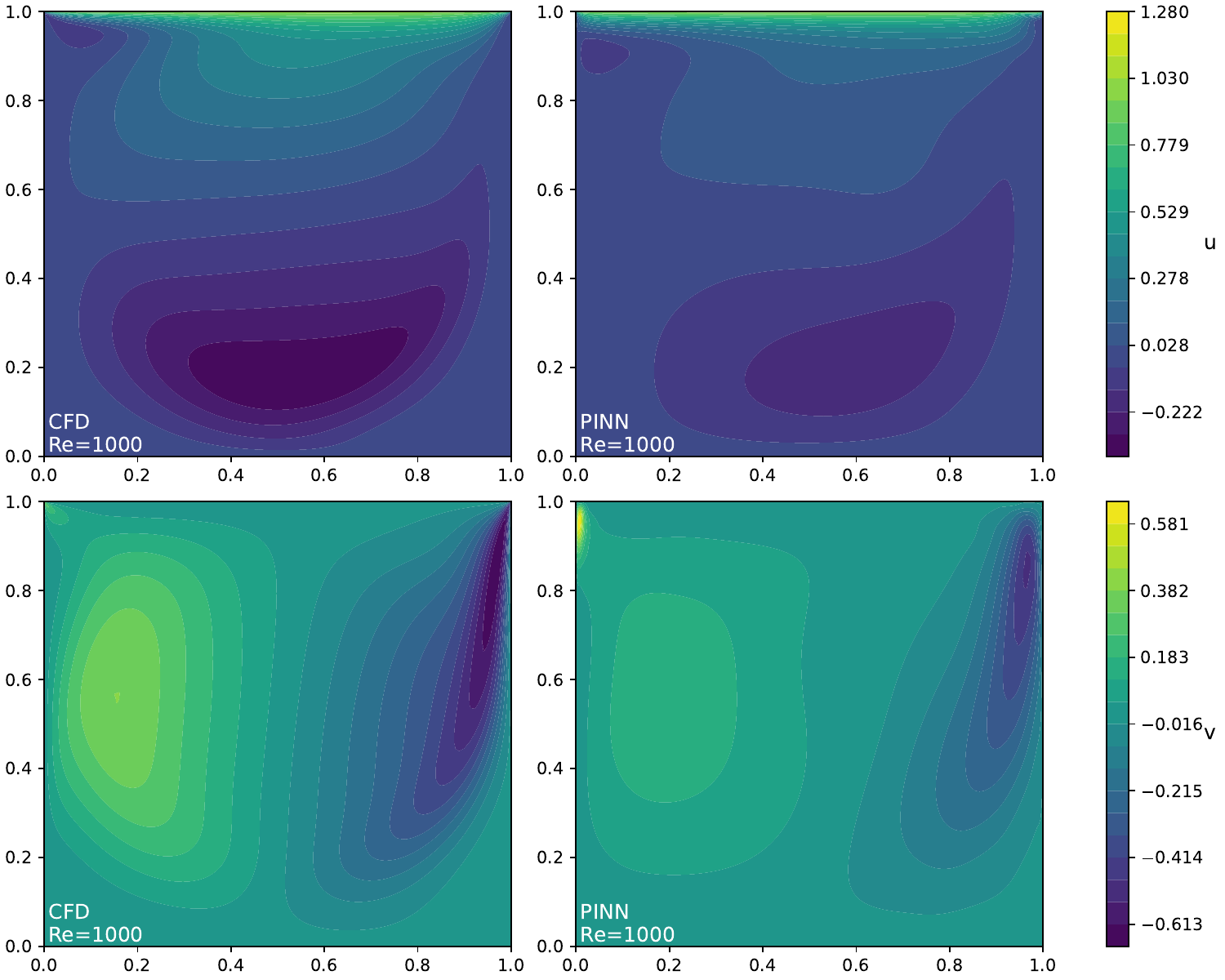}
        \label{fig:lid_driven_uvp_1000}
                \caption{Re=1000}
    \end{subfigure}  
    \caption{Comparison of the results obtained from PINN (right) and CFD simulations (left) for the lid-driven cavity at different Reynolds numbers: (a) Re =400 and (b) Re =1000. The top and bottom panels in each subfigure correspond to the $x$- and $y$-components of velocity ($u, v$).}
    \label{fig:lid_driven_uvp_all}
\end{figure}

We also present the velocity field results for intermediate and larger values of Reynolds Number ($Re = 400$ \& 1000) in Figure \ref{fig:lid_driven_vel_field_400_1000}. Similar to Figure \ref{fig:lid_driven_uvp}, the horizontal velocity field (u) and vertical velocity field (v) for $Re =$ 400 and 1000 are compared in Figure \ref{fig:lid_driven_uvp_all}a and \ref{fig:lid_driven_uvp_all}b respectively. 
It is observed that the position of the center of the primary vortex moves towards 
the geometric center of cavity as $Re$ increases. The comparison plots (Figure \ref{fig:lid_driven_vel_field_400_1000} and \ref{fig:lid_driven_uvp_all}) demonstrate that standard PINN framework begins to struggle in capturing the flow pattern at higher Reynolds Number where the fluid flow is more complex and dominated by inertial forces.

\begin{figure}[!htb]
    \centering
        \includegraphics[width=.9\textwidth]{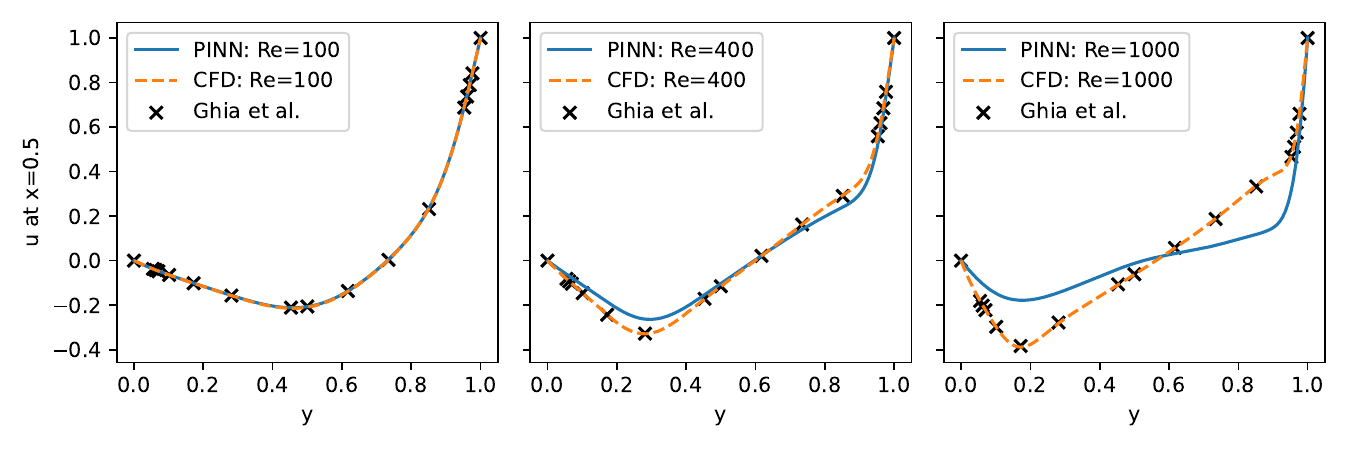}
        \includegraphics[width=0.9\textwidth]{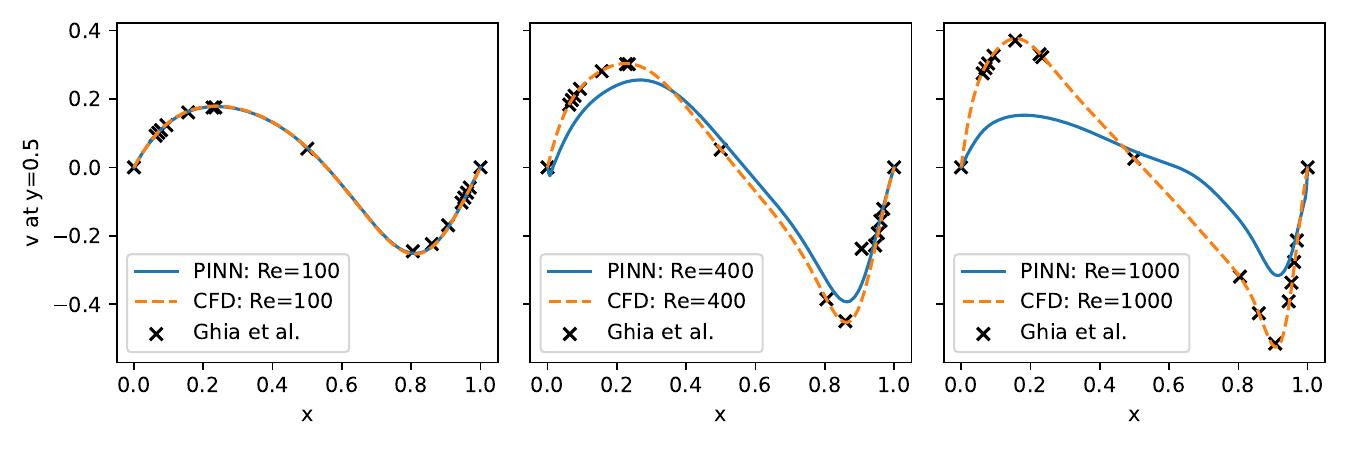}
    \caption{Velocity profiles along the vertical and horizontal centerlines of the cavity with $Re$ = 100 (left panel), 400 (middle panel) and 1000 (right panel). The solid yellow line, green dotted line and blue dotted points correspond to the results obtained from the PINN, CFD and Ghia et al \cite{GHIA1982387}.} 
    \label{fig:ghia_comp}
\end{figure}

To further illustrate the effectiveness of the PINN framework, we show 
the predicted $x$ and $y$ components of the velocity field along the vertical and horizontal centerlines of the cavity with different values of $Re$ in Figure \ref{fig:ghia_comp}. The top and bottom panels correspond to $u$ and $v$ respectively at $Re$ = 100 (left), 400 (middle) and 1000 (right). The blue solid and the green dotted lines denote the predictions from PINN and CFD simulations respectively. The black points represent the benchmark results of Ghia et al \cite{GHIA1982387}. At $Re=100$, the variations in $u$ and $v$ are in good agreement . 
At $Re=400$, the PINN prediction exhibits moderate deviation from our CFD simulation or Ghia et al \cite{GHIA1982387}, while at $Re=1000$ a significant discrepancy is observed. In particular, the PINN demonstrates less accuracy in capturing the extrema of $v$ compared to those of $u$ at $Re=1000$.
The $L_2$ normalized errors, total loss and numbers of iterations for different values of $Re$ are displayed in Table~\ref{tab:loss_lid}.
 The $L_2$ errors in the vertical velocity component, $(\epsilon_{L_2}^{(v)})$, 
 is larger than the horizontal component 
 ($\epsilon_{L_2}^{(u)}$) by a factor of $\sim 2$, which indicates that PINN 
 has captured the horizontal component of the velocity field more accurately.
 At $Re=100$, the $\epsilon_{L_2}^{(v)}$ error is  $\sim$ 11$\%$ and increased by a factor of $\sim 5$ at $Re=1000$. The total loss  is also larger by an order of $\sim 3$ for these two cases. This indicates a limitation of the PINN framework at higher Reynolds numbers.

\begin{table}[!htb]
\centering
\renewcommand{\arraystretch}{1.3}
\begin{tabular}{||c||c|c||c|c||}
\hline
\hline
\multirow{2}{*}{\textbf{Re}} & \multicolumn{2}{c||}{\textbf{$L_2$ normalized errors}} & \multicolumn{2}{c||}{\textbf{PINN}} \\
\cline{2-5}
 & $\epsilon_{L_2}^{(u)}$ & $\epsilon_{L_2}^{(v)}$ & Total Loss & No. of Iterations \\
\hline
100  & 0.0437 & 0.0810 & $3.77 \times 10^{-6}$ & 3,00{,}000 \\
\hline
400  & 0.1202 & 0.2369 & $1.55 \times 10^{-5}$ & 3,00{,}000 \\
\hline
1000  & 0.3203 & 0.5209 & $3.02 \times 10^{-3}$ & 3,00{,}000 \\
\hline
\hline
\end{tabular}
\caption{Error metrics, total loss, and iteration count for the lid-driven cavity flow}
\label{tab:loss_lid}
\end{table}

Figure \ref{fig:loss1} in Appendix~A, illustrates the training history of our PINN framework. We depict the 
 evolution of losses due to the $x$-momentum ($\mathcal{L}_{\text{mom}}^{(u)}$), $y$-momentum ($\mathcal{L}_{\text{mom}}^{(v)}$) and the continuity equation 
 ($\mathcal{L}_{\text{cont}}$) along with the total loss by varying the iteration number. 
Figure \ref{fig:loss1} shows that after $\sim 3 \times 10^5$ iterations, the convergence is achieved, and by the end of training, the total loss reaches an minimum value of $3.77 \times 10^{-6}$ with $Re = 100$.
Our network is able to capture the flow field with high accuracy at lower $Re$, primarily due to the application of Hard-BCs. Without this condition, the network would also need to learn the flow behavior at the boundaries, which would increase the model complexity and computational cost. The imposition of hard-boundary conditions has reduced the number of terms in the loss function, and the network is able to learn the flow field more efficiently. To demonstrate the advantage of Hard-BC enforcement, we study PINN models with both Hard and Soft-BC implementations at a fixed $Re$ and compare the results in Appendix~D. We observe that the most dominant contribution in total loss comes from the loss term due to boundary condition ($\mathcal{L}_{\text{Dirichlet}}$ or $\mathcal{L}_{\text{BC}}$) 
(see Figure \ref{fig:soft}) and the total loss differs by an order of $\sim 4$ compared to 
 $\mathcal{L}^{Hard-BC}$. Hardware details and training time are tabulated in 
 Table \ref{tab:training_times} in Appendix~B.


\subsection{Steady flow past a 2D circular obstacle in pipe}

\begin{figure}[!htb]
    \centering
    \includegraphics[width=0.69\textwidth]{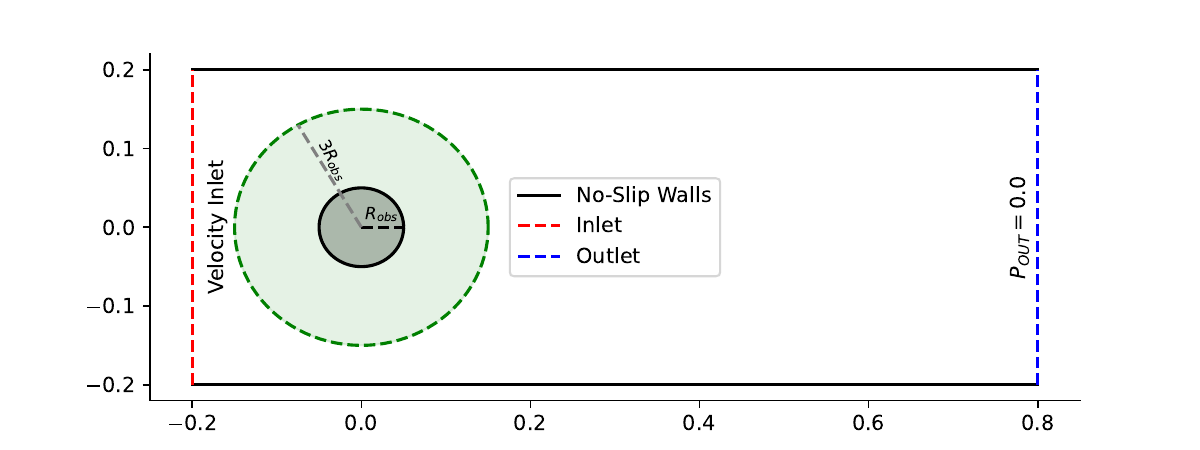}
    \caption{The geometry of the flow around a circular obstacle. 
 The radius of the circular obstacle is $R_{\text{obs}} = 0.05$ with the center placed at the origin, displayed as a black circle. 
 The green dotted circle has a radius of 3$R_{\text{obs}}$, where additional 
 collocations points are generated. }   
    \label{fig:circ_obs_geo}
\end{figure}

In this section, we study the steady-state flow past a 2D circular obstacle in a pipe as shown in Figure \ref{fig:circ_obs_geo} to validate our PINN framework. 
A circular obstacle with radius $R_{\text{obs}} = 0.05$ is placed at the origin, as shown in grey color in Figure \ref{fig:circ_obs_geo}. The top and bottom boundary walls, along with the obstacle boundary wall, are considered to be no-slip walls. At the outlet ($x = 0.8$), a zero pressure boundary condition is applied, and a parabolic velocity profile 
is assumed at $x = -0.2$ (inlet). The $x$ and $y$ components of inlet velocities 
are defined by 
\begin{equation}
    u_{\text{inlet}}(y) = \frac{4}{0.4^2} (0.2^2 - y^2), ~~~~ v_{\text{inlet}} = 0
\end{equation}
To impose the hard boundary conditions, the output functions in the  PINN framework are chosen as
\begin{eqnarray}
u(x, y) = u_{\text{NN}}(x, y) (r - R_{\text{obs}}) (y^2 - 0.2^2) (x + 0.2) + u_{\text{inlet}}(y) \frac{(r - R_{\text{obs}})}{\sqrt{y^2 + 0.2^2} - R_{\text{obs}}} \nonumber \\ 
        v(x, y) = v_{\text{NN}}(x, y) (r - R_{\text{obs}}) (0.2^2 - y^2) (x + 0.2)                 \label{eq:cir_hard} \\
                 p(x, y) = p_{\text{NN}}(x, y) (0.8 - x)   \nonumber
\end{eqnarray}
where $r = \sqrt{x^{2} + y^{2}}$. Similar to the lid-driven cavity scenario, we use three sub-neural networks to model 
the velocity and pressure fields ($u,v,p$). The collocation points are chosen as 1000 fixed random points in the domain $-0.2 \leq x \leq 0.8$ and $-0.2 \leq y \leq 0.2$, after which the points lying inside the obstacle are removed. We also choose 1000 additional random points in the region $R_{\text{obs}} \leq \sqrt{x^2 + y^2} \leq 3R_{\text{obs}}$ (see the green dotted circle in Figure \ref{fig:circ_obs_geo}) to better capture the flow features around the obstacle.
The complete list of hyperparameters is enlisted in Table~\ref{tab:hyperparameters_revised} in Sec~\ref{sec:hyper}.

\begin{figure}[!htb]
    \centering
        \includegraphics[width=0.99\textwidth]{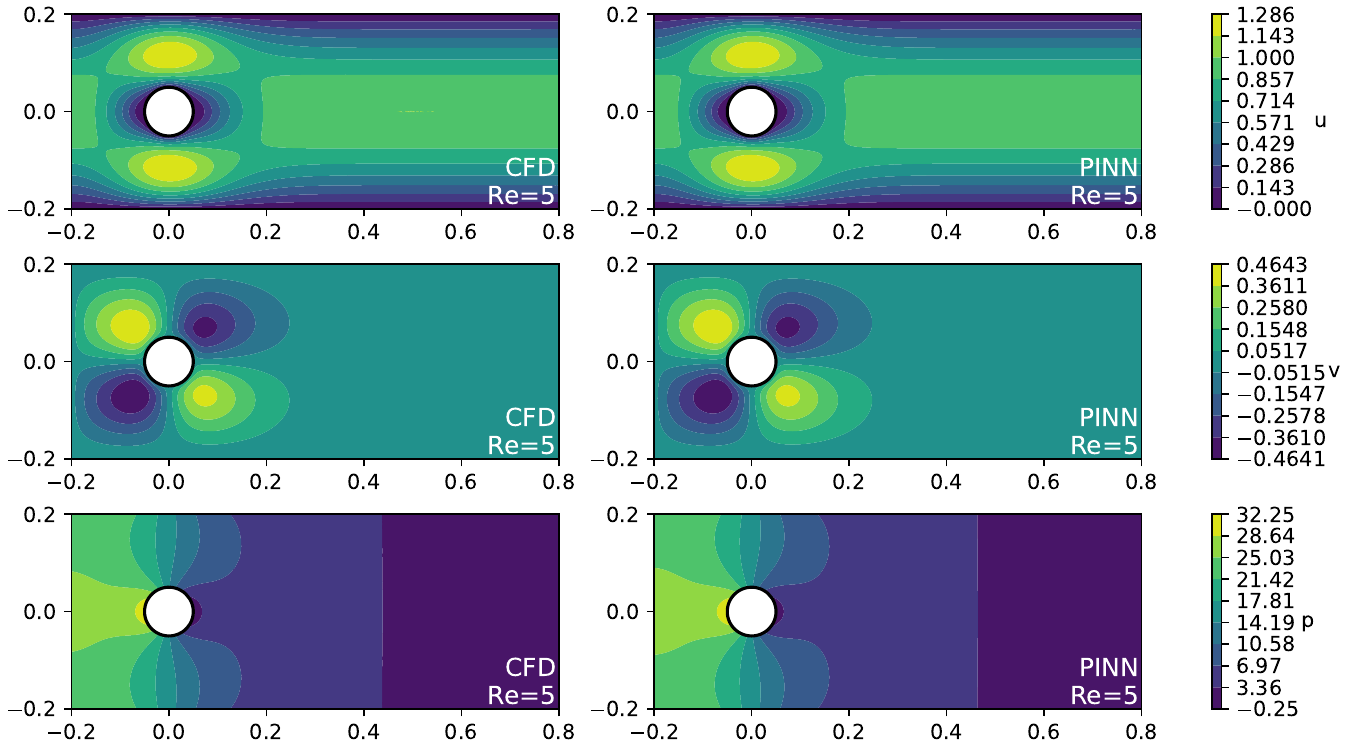}
            \caption{Comparison of the results obtained from CFD simulations (left) and PINN (right) frameworks for the flow past a circular obstacle at different Reynolds numbers. 
    In each case, the top, middle, and bottom panels correspond to $x$- and $y$-components of velocity ($u,v$), and pressure ($p$), respectively.}
            \label{fig:circ_obs_5}
\end{figure}

\begin{figure}[!htb]
    \begin{subfigure}[b]{0.48\textwidth}
        \centering
        \includegraphics[width=\textwidth]{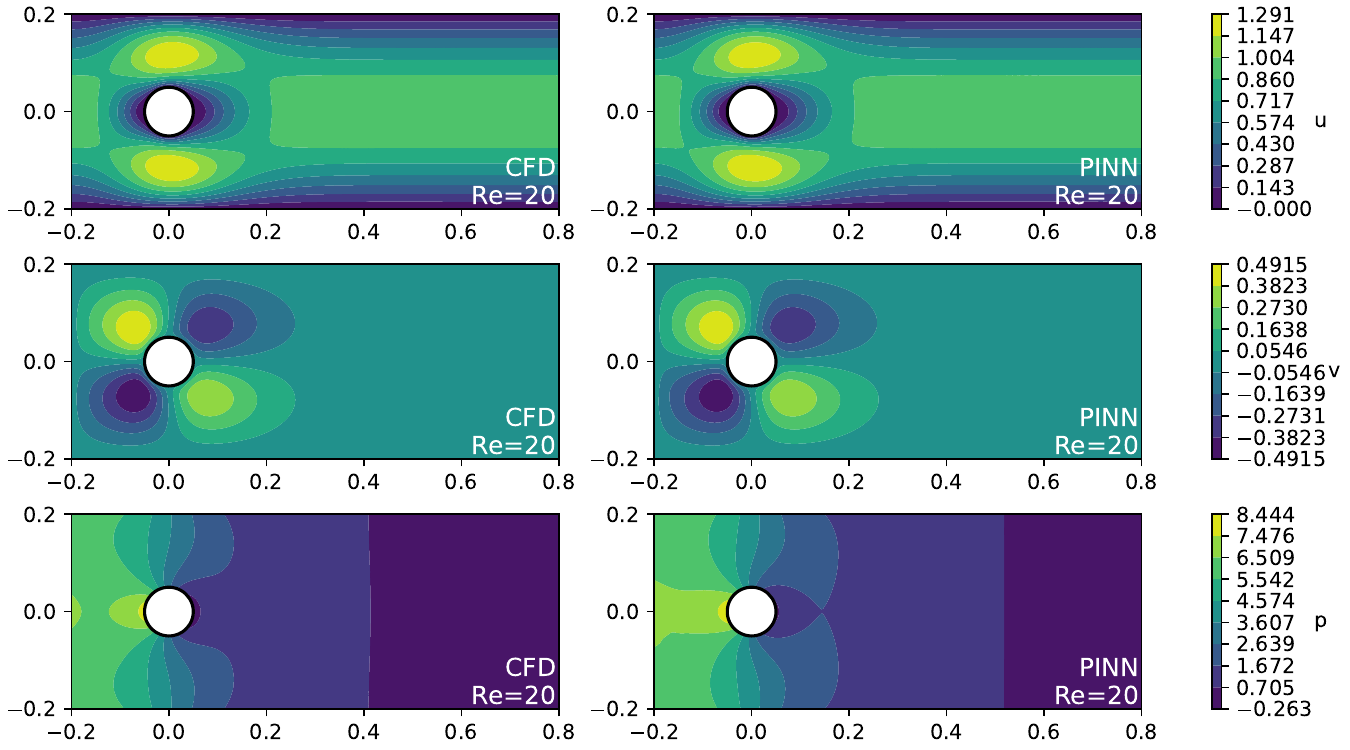}
        \label{fig:circ_obs_20}
                \caption{Re=20}
    \end{subfigure}    
    \begin{subfigure}[b]{0.48\textwidth}
        \centering
        \includegraphics[width=\textwidth]{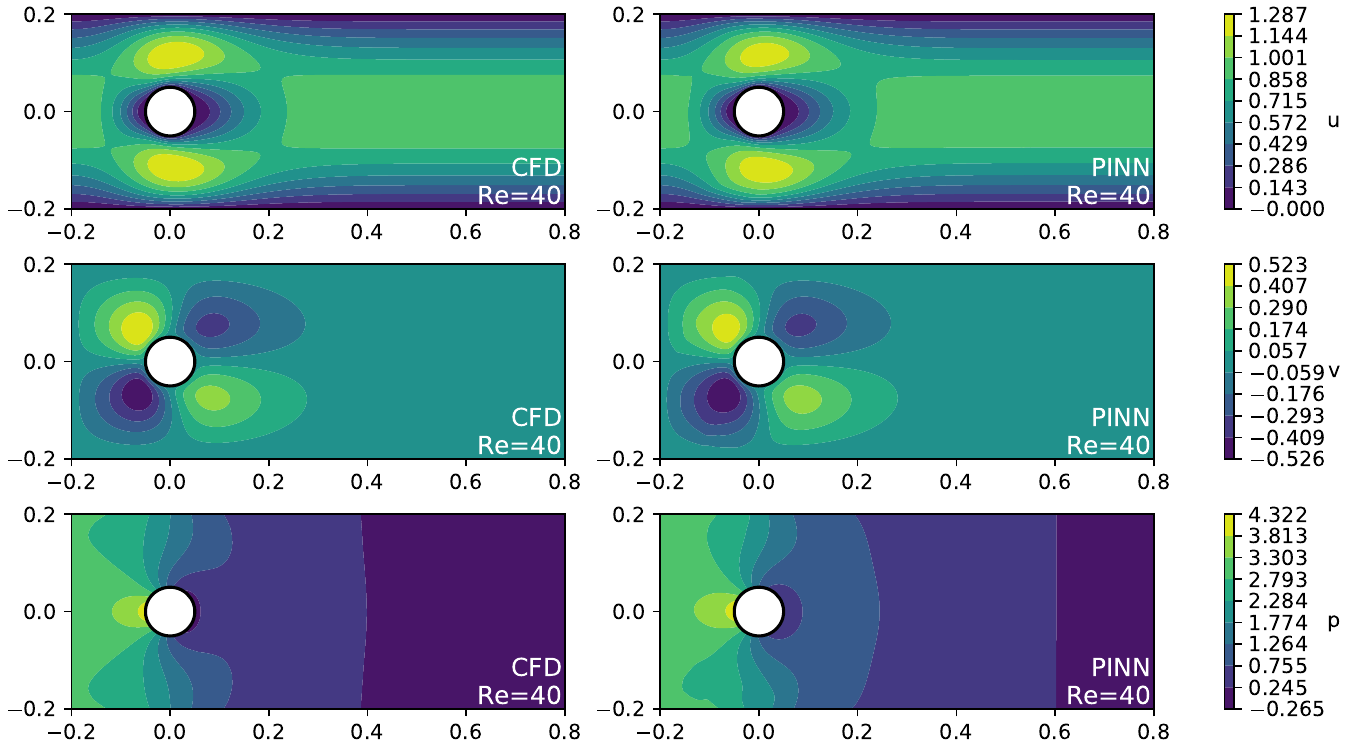}
        \label{fig:circ_obs_40}
                \caption{Re=40}
    \end{subfigure}  
    \caption{Comparison of the results obtained at (a)$Re$ = 20  and (b) $Re$ = 40 for the flow past a circular obstacle. In each case, the top, middle, and bottom panels correspond to $u,v$,$p$ respectively for CFD simulations (left) 
    and PINN (right) frameworks.}
    \label{fig:circ_obs_all}
\end{figure}

\begin{figure}[!htb]
    \centering
    \includegraphics[width=0.8\textwidth]{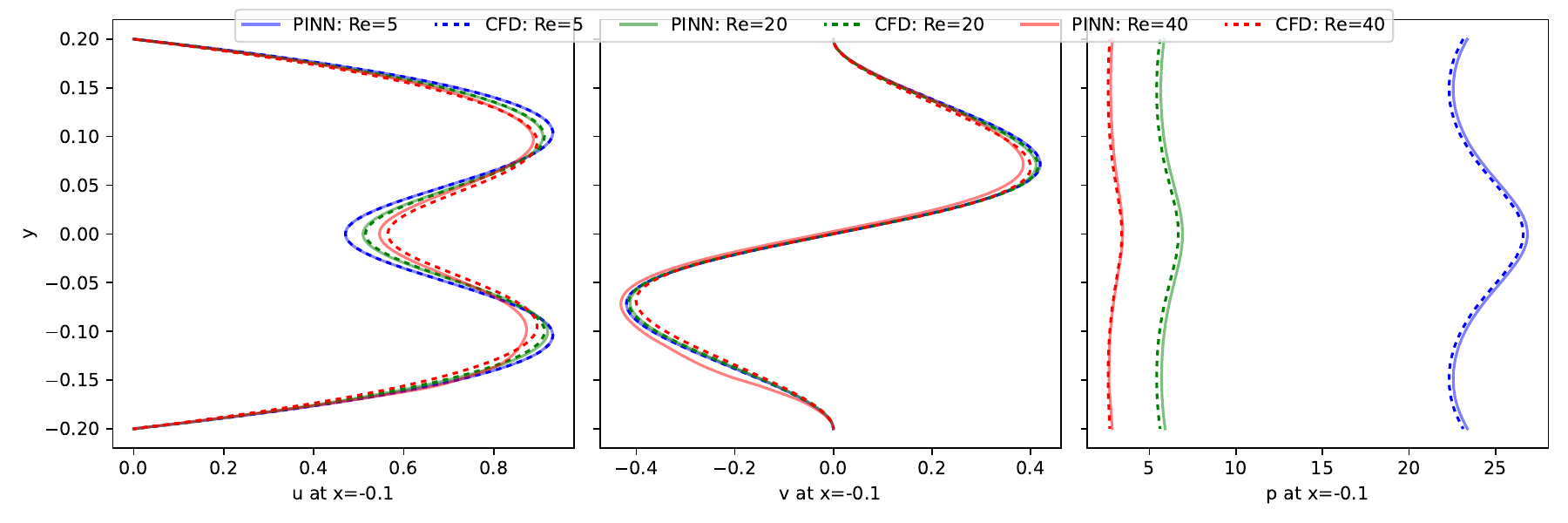}
        \includegraphics[width=0.8\textwidth]{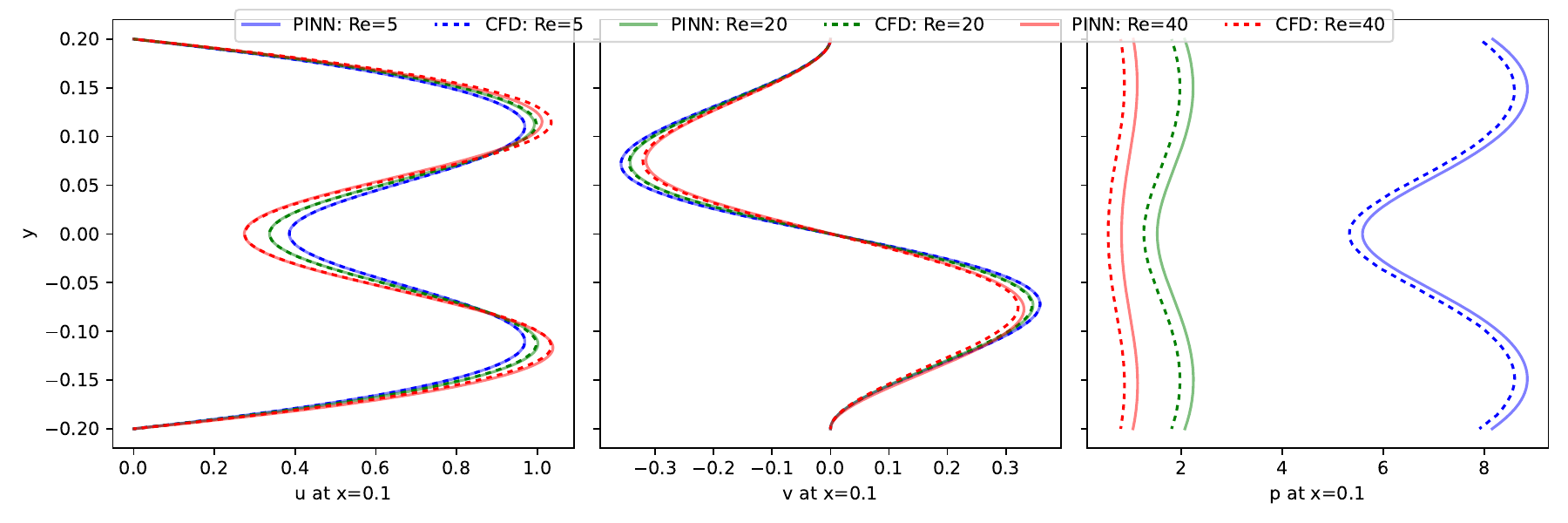}
    \caption{The comparisons of the velocity and pressure profiles at $x = -0.1$ (top panel) and at $x = 0.1$ (bottom panel). The center of the obstacle is located at $x=0.0$}
        \label{fig:circ_obs_uvp}
\end{figure}

\begin{figure}[!htb]
    \centering
    \includegraphics[width=0.6\textwidth]{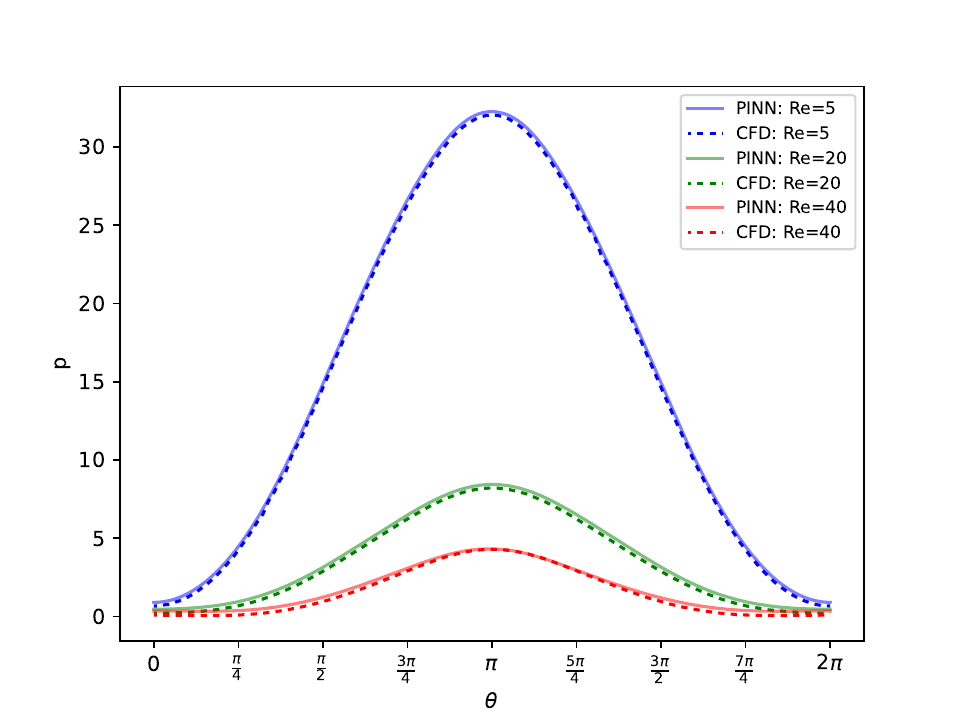}
    \caption{The variation of pressure on the surface of the circular obstacle as a function of angle $\theta$.}
            \label{fig:circ_obs_p}
\end{figure}

The comparison between the results obtained from the CFD simulation and the PINN for $Re = 5$ is shown in Figure \ref{fig:circ_obs_5} in the left and right columns, respectively. The top, middle, and bottom panels correspond to the $x$ and $y$ components of velocities and pressure, respectively. 
Figure \ref{fig:circ_obs_5} demonstrates that the proposed PINN framework effectively captures the key features of the flow. At the front of the circular obstacle, the fluid velocity approaches zero (as indicated by the dark blue region in the top panel). In this stagnation 
region, the pressure attains a maximum value (see the small dark yellow region in the bottom panel).
Due to the presence of the obstacle, the cross-sectional area is reduced.
The fluid velocity increases at the top and bottom of the obstacle, and pressure decreases. After the flow separation, we observe the wake formation, which is less chaotic due to the small Reynolds number we chose. 
We further consider two more cases with Reynolds number $Re = 20$ and 40, both of which are below the critical value for vortex shedding initiation. The CFD and PINN predicted velocity ($u,v$) and pressure ($p$) are presented in Figure\ref{fig:circ_obs_all}. The overall flow patterns remain similar to $Re = 5$ case. It should be noted that the maximum pressure decreases with increasing $Re$. 
We also compare the velocity and pressure profiles along the vertical lines at $x = -0.1$ (midway between the inlet and the obstacle) and $x = +0.1$ (downstream the obstacle) 
in the top and bottom panels of Figure\ref{fig:circ_obs_uvp}, respectively. Both the velocity and pressure profiles predicted by the PINN model exhibit good agreement with the CFD results (Figure\ref{fig:circ_obs_uvp}), whereas the PINN predicted pressure values are slightly higher in magnitude compared to those from CFD. The pressure distribution on the obstacle surface is crucial to quantify the drag forces and understand the flow features like 
flow separation, wake development etc. Figure\ref{fig:circ_obs_p} demonstrates that PINN predicted pressures on the surface of the obstacle (solid line) are also in good agreement with CFD simulations (dotted line) for different values of $Re$.

Drag coefficient ($C_d$), which is an integral quantity that quantifies the overall resistance exerted by the fluid on the obstacle, is defined as 
\begin{equation}
C_d = \frac{2 F_d}{ \rho U^2 A_{ref}}
\label{eq:force_drag}
\end{equation}
where $F_d$ is the total drag force, $\rho$ is fluid density, $U$ is the characteristic velocity and $A_{ref}$ is the reference area.
Pressure driven drag ($F_{dp}$) arises due to the pressure difference between 
the front and rear surfaces of the obstacle and the shear stresses acting tangentially to the surface generates viscous drag force ($F_{dv}$). Both these forces contribute to the total drag force and 
can be expressed as:
\begin{eqnarray}
F_{dp} = - \int_{0}^{2\pi} p(R,\theta)\,\cos\theta \; R \, d\theta
\label{eq:force_drag_p} \\
F_{dv} =  \int_{0}^{2\pi} \left[ \frac{2}{Re}\frac{\partial u}{\partial x}\cos\theta
+ \frac{1}{Re}\left(\frac{\partial u}{\partial y} + \frac{\partial v}{\partial x}\right)\sin\theta \right] R d\theta.
\label{eq:force_drag_v}
\end{eqnarray}
where $R$ is the radius of the obstacle and $\theta$ is the angle with respect to positive $x$ axis.  
\begin{table}[!htb]
\centering
\renewcommand{\arraystretch}{1.3}
\begin{tabular}{|c||c|c||c|c||c|c|}
\hline
\multirow{2}{*}{\textbf{Re}} & \multicolumn{2}{c||}{\textbf{Pressure Drag Coefficient }} & \multicolumn{2}{c||}{\textbf{Viscous Drag Coefficient }} & \multicolumn{2}{c|}{\textbf{Total Drag Coefficient}} \\
\multirow{2}{*}{} & \multicolumn{2}{c||}{\textbf{$C_{dp}$}} & \multicolumn{2}{c||}{\textbf{ $C_{dv}$}} & \multicolumn{2}{c|}{\textbf{$C_{d}$ = $C_{dp}$ + $C_{dv}$}} \\
\cline{2-7}
    & \textbf{CFD} & \textbf{PINN} & \textbf{CFD} & \textbf{PINN} & \textbf{CFD} & \textbf{PINN} \\
\hline
5 & 4.91 & 4.92 & 3.72 & 3.73 & 8.63 & 8.65 \\
\hline
20 & 1.24 & 1.24 & 0.94 & 0.93 & 2.18 & 2.17 \\
\hline
40 & 0.64 & 0.60 & 0.48 & 0.45 & 1.12 & 1.05 \\
\hline
\end{tabular}
\caption{$L_2$ normalized errors, total loss, and iteration count for the flow past a 2D circular obstacle.}
\label{tab:drag_coeff_cir}
\end{table}
Using Equations \ref{eq:force_drag_p}-\ref{eq:force_drag_v}, $C_{dp}$ and $C_{dv}$ can be evaluated and the total Drag coefficient ($C_d$) is the sum of these two 
coefficients ($C_d = C_{dp} + C_{dv}$). 
We present the pressure, viscous and total drag coefficient in Table \ref{tab:drag_coeff_cir} to demonstrate the performance of our PINN model across a range of $Re$ values. At the low Reynolds number, the flow remains 
laminar and the contributions from the viscous and pressure forces are of comparable 
magnitude. Overall, the PINN predicted drag coefficients values are in good agreement with those obtained from CFD, with the largest deviation is observed in $C_d$ at $Re =40$, with an error $\sim 6\%$.

\begin{table}[!htb]
\centering
\renewcommand{\arraystretch}{1.3}
\begin{tabular}{|c||c|c|c||c|c|}
\hline
\multirow{2}{*}{\textbf{Re}} & \multicolumn{3}{c||}{\textbf{$L_2$ normalized errors}} & \multicolumn{2}{c|}{\textbf{PINN}} \\
\cline{2-6}
 & $\epsilon_{L_2}^{(u)}$ & $\epsilon_{L_2}^{(v)}$ & $\epsilon_{L_2}^{(p)}$ & Total Loss & No. of Iterations \\
\hline
5 & 0.0006 & 0.0011 & 0.0224 & $1.3 \times 10^{-3}$ & 3{,}00{,}000 \\
\hline
20 & 0.0032 & 0.0158 & 0.0975 & $4.5 \times 10^{-4}$ & 3{,}00{,}000 \\
\hline
40 & 0.0234 & 0.0790 & 0.1886 & $8.8 \times 10^{-4}$ & 3{,}00{,}000 \\
\hline
\end{tabular}
\caption{$L_2$ normalized errors, total loss, and iteration count for the flow past a 2D circular obstacle.}
\label{tab:loss_cir}
\end{table}

The $L_2$ normalized errors and the total loss are presented in Table~\ref{tab:loss_cir}, where it can be seen that that the effectiveness of PINN decreases with increasing  $Re$ values. For instance, $\epsilon_{L_2}^{(u)}$ increases from $\mathcal{O}$($10^{-4}$) to $\mathcal{O}$($10^{-2}$) as $Re$ value changes from 5 to 40. As noted earlier, the largest deviation occurs in the pressure profile. The minimum $L_2$ normalized error in pressure is $\sim$ 0.02 at $Re=5$ and $\epsilon_{L_2}^{(p)}$ increases up to $\sim$ 0.19 at $Re=40$. 
It is worth mentioning that Rao et al. \cite{RAO2020207} reported a $L_2$ normalized error $\mathcal{O}$($10^{-2}$) for the velocity fields for the flow past a 2D circular obstacle at $Re=5$. We find that the employment of hard-boundary conditions has improved the error by \textit{two orders of magnitude}.
We also present the training history for different loss terms in 
Figure \ref{fig:loss2}.

\subsection{Steady flow in a pipe with sinusoidal boundary}
\label{sec:sine_pipe_steady}

\begin{figure} [!h]
    \centering
    \includegraphics[width=0.8\textwidth]{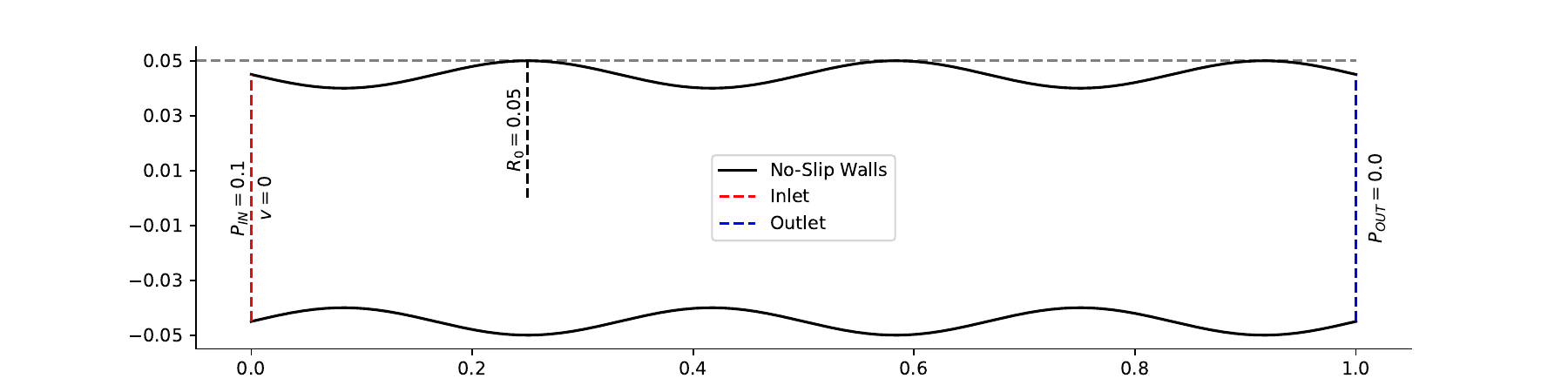}
    \caption{The geometry of the 2D pipe flow with sinusoidal boundary. Here $R_0 = 0.05$, $A = 0.005$, $N = 6$ and $L=$ 1. The inlet and outlet pressures are fixed as 0.1 at $x = 0$ and 0.0 at $x = 1$, respectively.}
    \label{fig:sin_bound_geo}
\end{figure}

In this section, we explore the flow characteristics in a sinusoidal channel. Flows in a tube or channel with flat and wavy walls play a significant role in various industrial applications, natural and physiological phenomena like conduit flow, stenotic and aneurysmal blood flow, urinary flow, etc. \cite{berger2000flows, brisman2006cerebral, ram2023motion}. Mostly, these are pressure-driven flows, and realistic flow geometries (e.g., vascular) are often very complex and irregular. We consider a symmetric pipe flow with a sinusoidal boundary, and the radius of the channel is defined as:
\begin{equation}
 R(x) = R_{0} - A(1 + \sin(N \pi x/L))
\end{equation}
where $L$, $A$ and $N$ are the length of the pipe, the amplitude and wave number at the boundary. For the illustration purpose, we fix the parameters as $R_0 = 0.05$, $A = 0.005$, $N = 6$ and $L=1$. The inlet and outlet pressure difference govern the flow in the channel. The boundary conditions for pressure are chosen as: $P_{\text{in}} = $ 0.1 at the inlet ($x = 0$) and $P_{\text{out}} = $ 0.0 at the outlet ($x = 1$). The geometry of the channel and the boundary conditions are displayed in Figure \ref{fig:sin_bound_geo}. No slip conditions are also applied at the wall boundary.

\begin{figure}[!htb]
    \centering
        \includegraphics[width=0.9\textwidth]{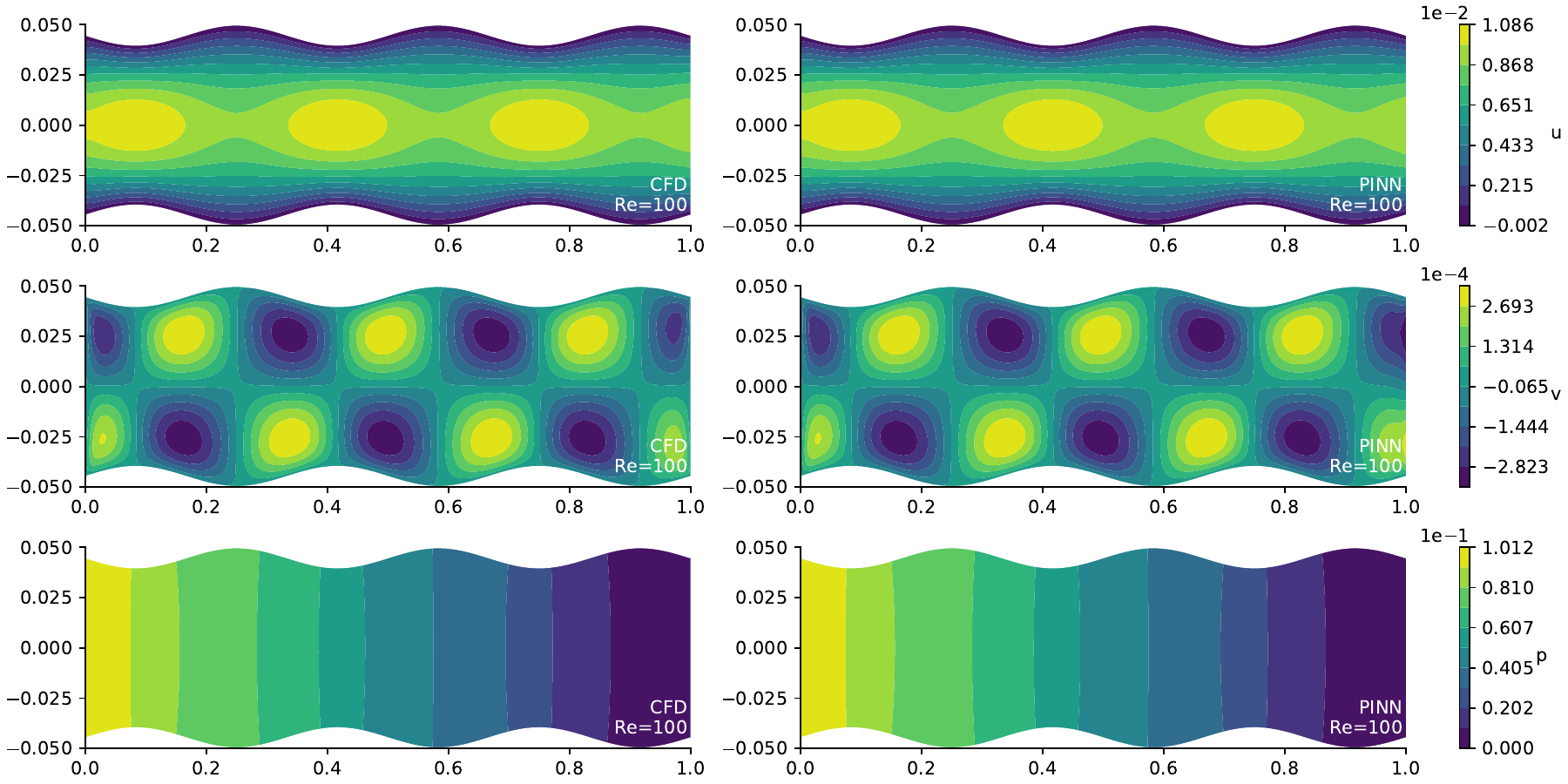}
                \label{fig:sin_bound_100}
                \includegraphics[width=0.9\textwidth]{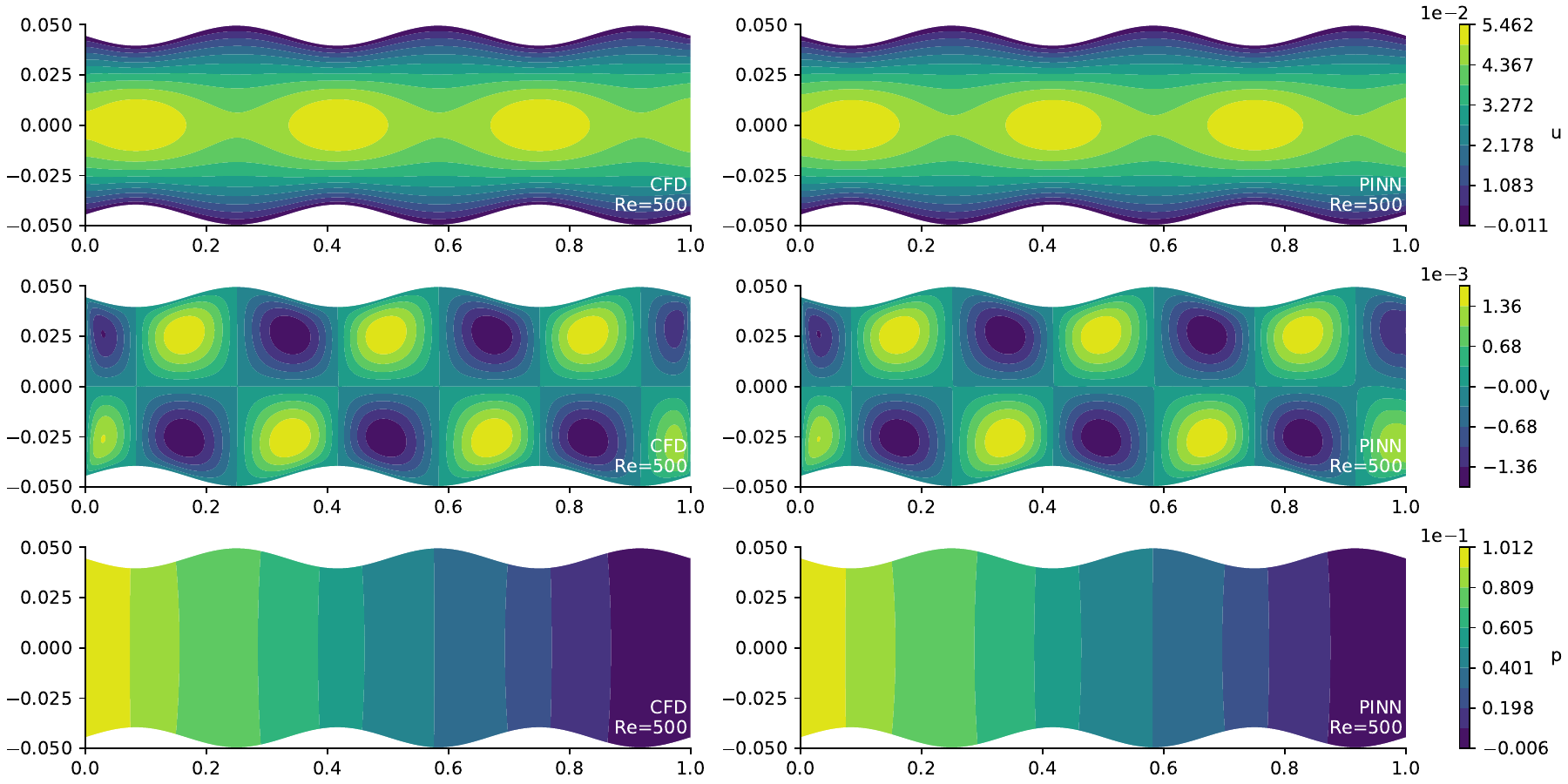}
                        \label{fig:sin_bound_500}
    \caption{Comparison of the results obtained from CFD simulations (left) and PINN framework (right) for the sinusoidal channel with Re = 100 and 500.}
    \label{fig:sin_bound_all}
\end{figure}

To enforce these boundary conditions in the \texttt{hard} way, the PINN model outputs are constructed as follows: 
\begin{equation}
    \begin{split}
 u(x,y) &= u_{\text{NN}}(x,y) \{ R(x)^{2} - y^{2} \} \\
 v(x,y) &= v_{\text{NN}}(x,y) x \{ R(x)^{2} - y^{2} \} \\
 p(x,y) &= p_{\text{NN}}(x,y) x(1 - x) + P_{\text{in}}(1 - x) + P_{\text{out}}x,
    \end{split}
    \label{eq:sine_hard}
\end{equation}
We have chosen three feed-forward neural networks with three hidden layers, each of 64 neurons. At every iteration we chose 1000 random collocation points in the domain $0 \leq x \leq 1$ and $-0.05 \leq y \leq 0.05$, after which the points lying outside the channel are removed. The complete list of hyperparameters are enlisted in Table~\ref{tab:hyperparameters_revised} in Sec~\ref{sec:hyper}.

\begin{figure}[!htb]
    \centering
    \begin{subfigure}[b]{0.49\textwidth}
        \includegraphics[width=\textwidth]{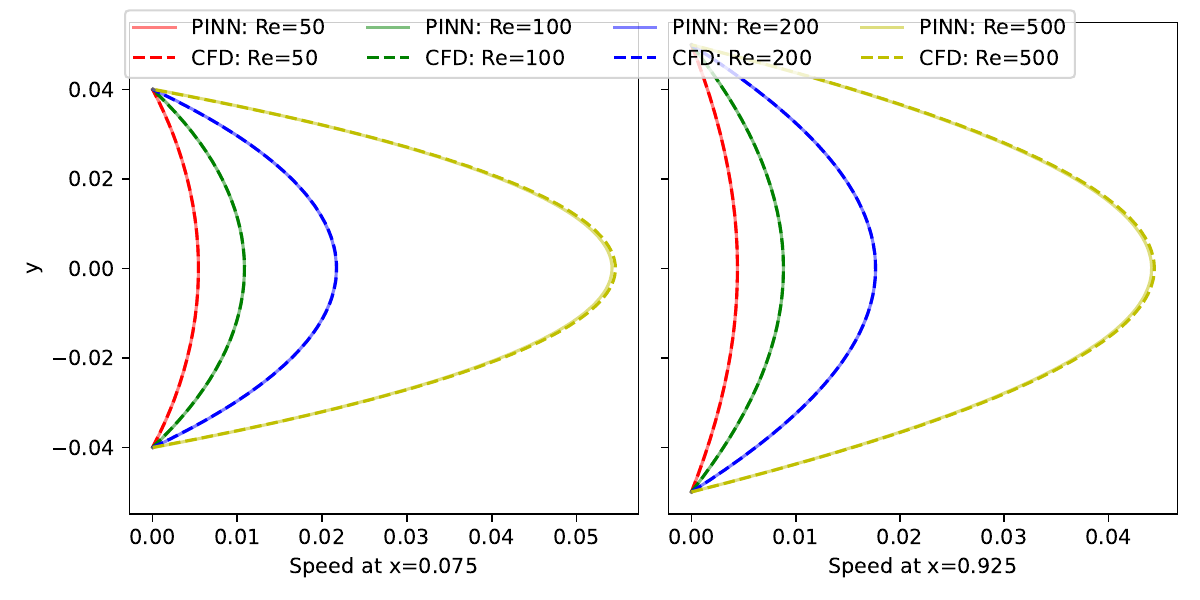}
        \caption{Velocity magnitude at $x$ = 0.075 and 0.925.}
        \label{fig:sin_bound_speed}
    \end{subfigure}
    \begin{subfigure}[b]{0.49\textwidth}
        \includegraphics[width=\textwidth]{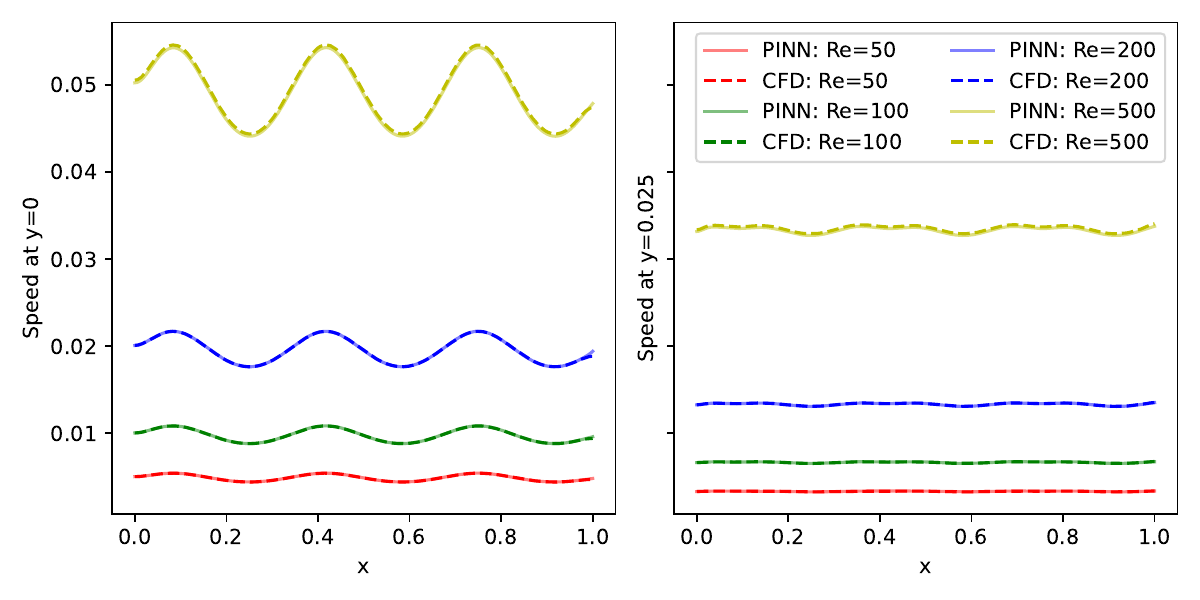}
        \caption{Velocity magnitude at $y$ = 0.0 and 0.025.}
        \label{fig:sin_bound_speed_y}
    \end{subfigure}
    \caption{
    The comparisons of the velocity magnitude (a) at different horizontal locations 
         ($x$ =0.075 and 0.925), and (b) at different vertical locations  ( at $y$ = 0.0 and 0.025) of the sinusoidal channel. The solid line and dotted line correspond to the results obtained from the PINN and CFD, respectively. Red, green, blue, yellow colors represent the results corresponding to $Re = $ 50, 100, 200 and 500. }
    \label{fig:sin_bound}
\end{figure}

We also study the flow pattern, velocity and pressure profiles for this 2D pipe at different Reynolds number: $Re$ = 50, 100, 200, and 500. 
For illustration purpose, we present the PINN predicted flow patterns at $Re = 100$ and 500 in top and bottom three rows in Figure \ref{fig:sin_bound_all} right column.
The top, middle and bottom panels for each $Re$ value correspond to the $x$ and $y$ components of velocity fields and pressure, respectively. The corresponding CFD results are presented in the left column of Figure \ref{fig:sin_bound_all}. 
It is observed that flow is predominantly governed by the streamwise (horizontal) velocity component ($u$), e.g., the maximum magnitude of the spanwise velocity component ($v$) is nearly two orders smaller than that of $u$. Fluid is accelerated at the narrower part of the channel due to the decrease in cross-sectional area and $u$ becomes maximum at $R = R_{minimum}$. The flow subsequently decelerates and slows down in the expanded region. As illustrated in Figure \ref{fig:sin_bound}, PINN predicted $u,v,p$ field contours show good agreement with the CFD simulations for the chosen $Re$ values. For further illustration of our PINN accuracy, in Figure \ref{fig:sin_bound_speed} we present the velocity magnitude along the vertical lines at the crest ($x$ =0.075) and trough ($x$ =0.925), which are located closest to the inlet and outlet respectively. 
Figure \ref{fig:sin_bound_speed} shows that the velocity profile is parabolic and the maximum velocity is obtained at the center of pipe. Our PINN predicted velocity reproduces both the profile shape and magnitude satisfactorily for $Re = $ 50, 100, 200 and 500. We also present the velocity profiles along the horizontal lines at $y$ =0.0 and 0.025 in Figure \ref{fig:sin_bound_speed_y} for the same $Re$ values. The PINN predictions closely match with the CFD results.

$L_2$-norm errors and total loss for $Re = $ 50, 100, 200 and 500 are tabulated in Table~\ref{tab:loss_sinu}. 
Maximum $L_2$ normalized error is observed for $v$ ($\mathcal{O}(10^{-2})$), 
whereas $\epsilon_{L_2}^{(u)}$ ($\mathcal{O}(10^{-3})$) and $\epsilon_{L_2}^{(p)}$ ($\mathcal{O}(10^{-4})$) are relatively smaller.
We also present the training history for different loss terms in
Figure \ref{fig:loss3} in Appendix~A.
After $3 \times 10^{5}$ iterations, the minimum total loss obtained is $1.04 \times 10^{-7}$ and $1.1 \times 10^{-6}$ at Re = 50 and 500 respectively.

\begin{table}[h!]
\centering
\renewcommand{\arraystretch}{1.3}
\begin{tabular}{|c||c|c|c||c|c|}
\hline
\multirow{2}{*}{\textbf{Re}} & \multicolumn{3}{c||}{\textbf{$L_2$ normalized errors}} & \multicolumn{2}{c|}{\textbf{PINN}} \\
\cline{2-6}
 & $\epsilon_{L_2}^{(u)}$ & $\epsilon_{L_2}^{(v)}$ & $\epsilon_{L_2}^{(p)}$ & Total Loss & No. of Iterations \\
\hline
50 & 0.0019 & 0.0862 & 0.0008 & $1.04 \times 10^{-7}$ & 300{,}000 \\
\hline
100 & 0.0016 & 0.0811 & 0.0007 & $1.67 \times 10^{-7}$ & 300{,}000 \\
\hline
200 & 0.0020 & 0.1082 & 0.0005 & $2.25 \times 10^{-7}$ & 300{,}000 \\
\hline
500 & 0.0061 & 0.0571 & 0.0063 & $1.10 \times 10^{-6}$ & 300{,}000 \\
\hline
\end{tabular}
\caption{$L_2$ normalized errors, total loss, and iteration count for the flow in a pipe with a sinusoidal boundary.}
\label{tab:loss_sinu}
\end{table}

\subsection{Transient flow in a pipe with sinusoidal boundary}

In this section, we investigate the transient flow patterns in a pipe with sinusoidal boundary using the same geometry as described in previous subsection (Sec \ref{sec:sine_pipe_steady}). In the transient case, the temporal variations of the velocity fields are present and the non-dimensional NSEs as 
mentioned in Equation \ref{eqn:nse_non_dim1} -\ref{eqn:nse_non_dim3} take the form: 
\begin{eqnarray}
\frac {\partial u}{\partial x} +  \frac{\partial v}{\partial y} =  0 
				  \label{eqn:nse_non_dim4}			  \\
\frac {\partial u}{\partial t} + u \frac {\partial u}{\partial x} + v \frac {\partial u}{\partial y}  = 
- 	\frac {\partial p}{\partial x} + 
	\frac {1}{Re} 	(\frac {\partial^2 u}{\partial x^2}  
		+ \frac {\partial^2 u}{\partial y^2}  )  \\
  \frac {\partial v}{\partial t}  + u \frac {\partial v}{\partial x} + v \frac {\partial v}{\partial y}  = 
- \frac {\partial p}{\partial y} + 
	\frac{1}{Re} 	(\frac {\partial^2 v}{\partial x^2}  
		+ \frac {\partial^2 v}{\partial y^2} )  
				  \label{eqn:nse_non_dim5}
				  \end{eqnarray}
Correspondingly, the loss terms defined in Equation \ref{eqn:loss_hard} are modified to:
\begin{eqnarray}
    \mathcal{L}_{\text{mom}}^{(u)} = \frac{1}{|\mathcal{X}|} \sum_{i \in \mathcal{X}} \left[ \frac{\partial u}{\partial t} + u \frac{\partial u}{\partial x} + v \frac{\partial u}{\partial y} + \frac{\partial p}{\partial x} - \frac{1}{\text{Re}} \left( \frac{\partial^2 u}{\partial x^2} + \frac{\partial^2 u}{\partial y^2} \right) \right]^2_{i} \nonumber \\
    \mathcal{L}_{\text{mom}}^{(v)} = \frac{1}{|\mathcal{X}|} \sum_{i \in \mathcal{X}} \left[ \frac {\partial v}{\partial t}  + u \frac{\partial v}{\partial x} + v \frac{\partial v}{\partial y} + \frac{\partial p}{\partial y} - \frac{1}{\text{Re}} \left( \frac{\partial^2 v}{\partial x^2} + \frac{\partial^2 v}{\partial y^2} \right) \right]^2_{i} \\
    \mathcal{L}_{\text{cont}} = \frac{1}{|\mathcal{X}|} \sum_{i \in \mathcal{X}} \left[ \frac{\partial u}{\partial x} + \frac{\partial v}{\partial y} \right]^2_{i} \nonumber
   				  \label{eqn:loss_hard_time}
\end{eqnarray}

We assume that the flow is initially still. Instead of applying a constant inlet pressure, we impose a time-dependent inlet pressure given by:
\begin{eqnarray}
P_{in}(t) = \frac{1}{2} \left[ 1-cos\left(\frac{2\pi t}{T}\right) \right] P_{max} =  \frac{1}{2} \left[ 1 + Sin\left(\frac{2\pi t}{T} + \frac{3\pi}{2}\right) \right] P_{max}
\end{eqnarray}
where $P_{max}$ is the maximum pressure and set to 0.1 (similar to steady state case) and the period $T$ is taken as 1.0. 
\begin{figure}[!htb]
    \centering
        \includegraphics[width=\textwidth]{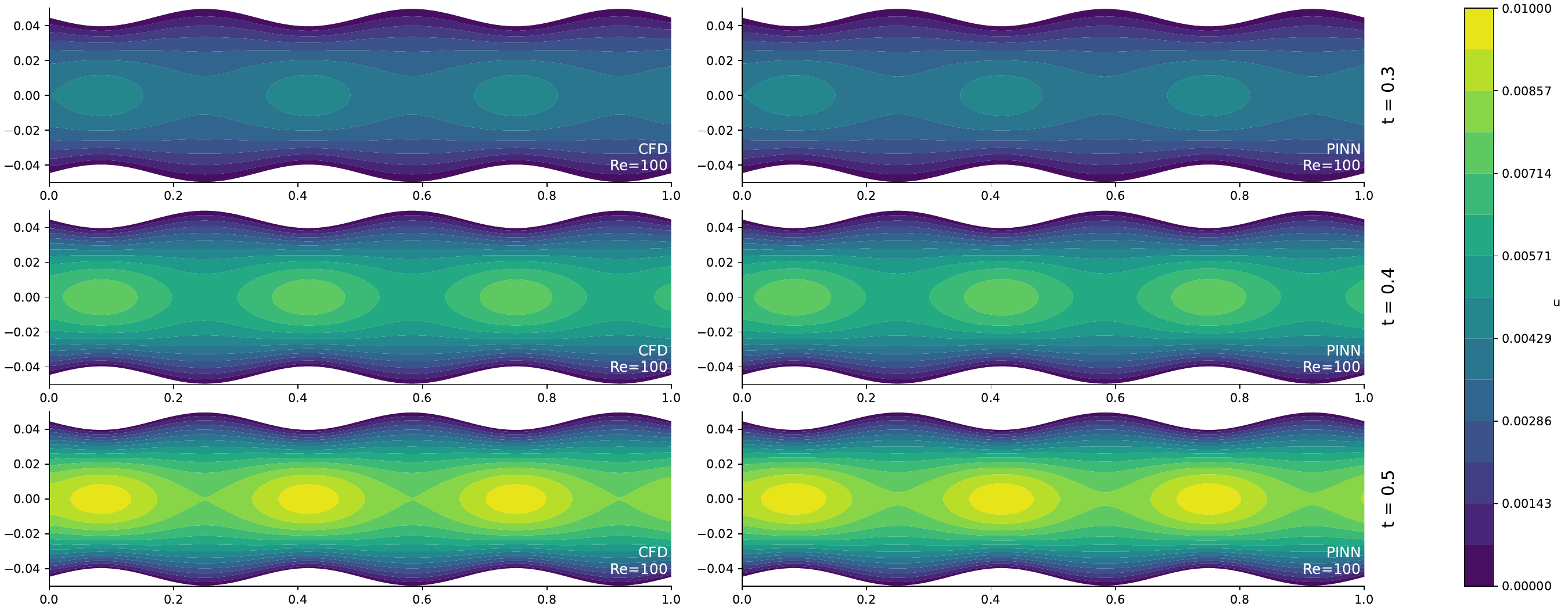}
    \caption{Comparison of the $x$ component of velocity field ($u$) obtained from CFD simulations (left) and PINN framework (right) for the transient flow in sinusoidal channel with Re = 100. The top, middle and bottom rows correspond to t = 0.3, 0.4 and 0.5.}
                \label{fig:sin_bound_100_td_u}
\end{figure}
Rest of the boundary conditions are identical to those as mentioned in Sec \ref{sec:sine_pipe_steady} and illustrated in Figure \ref{fig:sin_bound_geo}. To 
enforce these boundary conditions in the \texttt{hard} way, the PINN model outputs are constructed as follows: 
\begin{equation}
    \begin{split}
 u(x,y,t) &= u_{\text{NN}}(x,y,t) \{ R(x)^{2} - y^{2} \}t \\
 v(x,y,t) &= v_{\text{NN}}(x,y,t) x \{ R(x)^{2} - y^{2} \}t \\
 p(x,y,t) &= p_{\text{NN}}(x,y,t) x(1 - x) + P_{\text{in}}(t)(1 - x) + P_{\text{out}}x,
    \end{split}
    \label{eq:sine_hard_td}
\end{equation}
To 
enforce the initial conditions, we have multiplied $u$ and $v$ by $t$, which ensures that both $u$ and $v$ are zero at $t=0$. We choose 10,000 fixed collocation points randomly sampled in the spatio-temporal domain $0 \leq x \leq 1$, $-0.05 \leq y \leq 0.05$ and $0 \leq t \leq 1$, after which the points lying outside the channel are removed. The complete list of hyperparameters is enlisted in Table~\ref{tab:hyperparameters_revised} in Sec~\ref{sec:hyper}.

\begin{figure}[!htb]
    \centering
            \includegraphics[width=\textwidth]{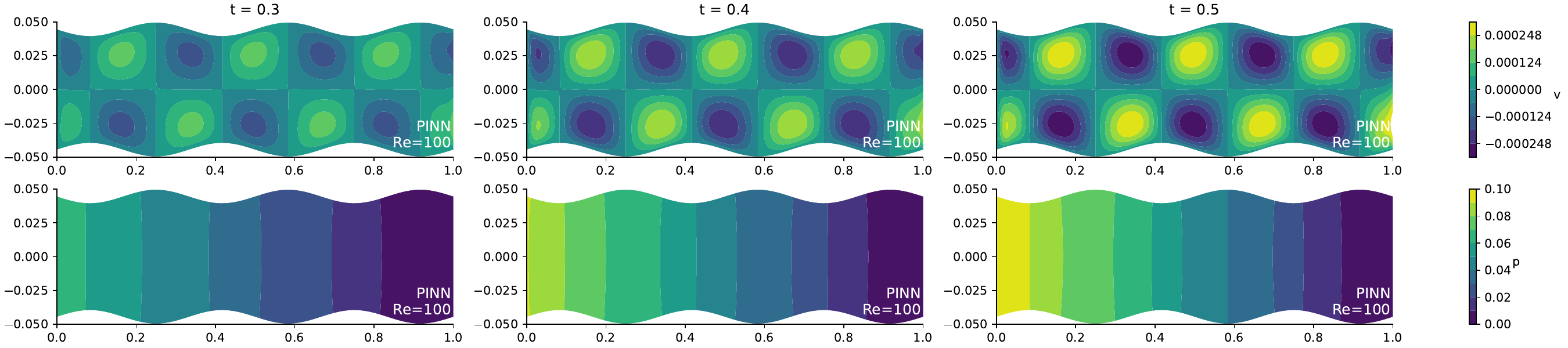}
    \caption{Y-component velocity field (top panel) and pressure field (bottom) predicted by the PINN framework for the transient flow in sinusoidal channel with Re = 100 at time t = 0.3, 0.4 and 0.5.}
                \label{fig:sin_bound_100_td_vp}
\end{figure}

\begin{figure}[!htb]
    \centering
    \begin{subfigure}[b]{0.49\textwidth}
        \includegraphics[width=\textwidth]{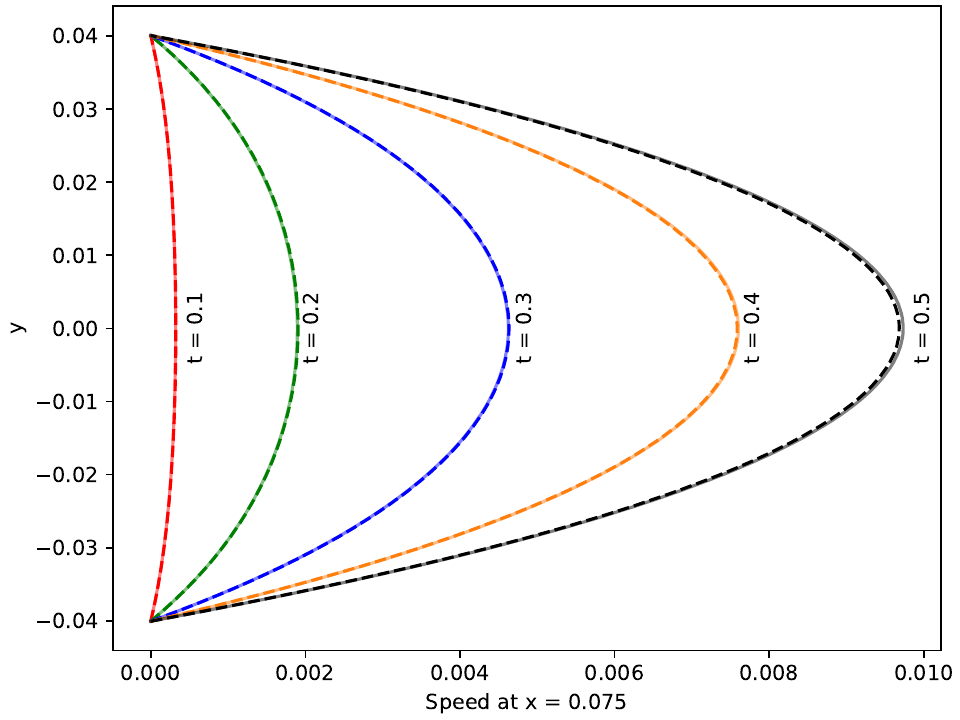}
        \label{fig:sin_bound_speed_x_td}
    \end{subfigure}
    \begin{subfigure}[b]{0.49\textwidth}
        \includegraphics[width=\textwidth]{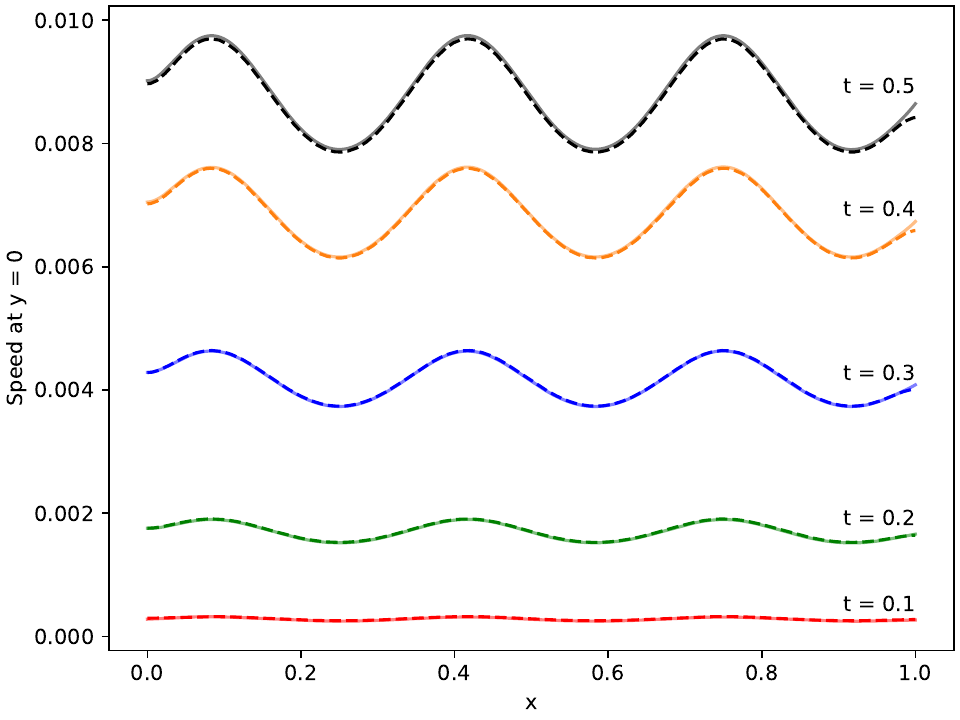}
        \label{fig:sin_bound_speed_y_td}
    \end{subfigure}
    \caption{ The comparisons of the transient velocity magnitude profiles (a) along vertical line at $x$ =0.075, and (b) along horizontal center-line (at $y$ = 0.0) of the sinusoidal channel with Re =100 at different times. The solid line and dotted line correspond to the results obtained from the PINN and CFD, respectively. Red, green, blue, yellow and black color represents the results corresponding to $t = $ 0.1, 0.2, 0.3, 0.4 and 0.5.}
        \label{fig:sin_bound_speed_td}
\end{figure}

\begin{table}[h!]
\centering
\renewcommand{\arraystretch}{1.3}
\begin{tabular}{|c||c|c|c||c|c|}
\hline
\hline
\multirow{2}{*}{\textbf{Time (s)}} & \multicolumn{3}{c||}{\textbf{$L_2$ normalized errors}} & \multicolumn{2}{c|}{\textbf{PINN}} \\
\cline{2-6}
 & $\epsilon_{L_2}^{(u)}$ & $\epsilon_{L_2}^{(v)}$ & $\epsilon_{L_2}^{(p)}$ & Total Loss & No. of Iterations \\
\hline
0.1 & 0.0216 & 0.1638 & 0.002 & \multirow{4}{*}{8.74$\times 10^{-8}$} & \multirow{4}{*}{3.0$\times 10^{5}$} \\
\cline{1-4}
0.2 & 0.005 & 0.0791 & 0.004 &  & \\
\cline{1-4}
0.3 & 0.0025 & 0.0882 & 0.0044 & & \\
\cline{1-4}
0.4 & 0.0035 & 0.0871 & 0.002 &  & \\
\cline{1-4}
0.5 & 0.0058 & 0.1003 & 0.0008 & & \\
\hline
\hline
\end{tabular}
\caption{$L_2$ normalized errors, total loss, and iteration count for the transient flow in a pipe with a sinusoidal boundary at $Re$ = 100.}
\label{tab:loss_sinu_td}
\end{table}

Figure \ref{fig:sin_bound_100_td_u} compares the results of the time-dependent studies done at Re=100 using CFD and PINN. Right side panel displays the progressive time evolution (t=0.3, 0.4, 0.5) of x-velocity, found in our PINN models. The increasing time steps show increase in velocity magnitude of the system, however the flow pattern remains same. 
When compared, left panel results from corresponding CFD simulations also evolve in similar manner with time. Outcomes of PINN model for the transient flow velocity in y direction and pressure are displayed in Figure \ref{fig:sin_bound_100_td_vp}. 
Time progression from t=0.3 to t= 0.5 causes rise in both y-velocity and pressure indicative of approaching a steady flow. 
The change in velocity magnitude is quantitatively analysed in Figure \ref{fig:sin_bound_speed_td} along $y$ and $x$ direction at $x$ =0.075 and at $y$ = 0.0 respectively. It is also evident here that the speed in both directions increases with time, keeping the nature of flow distribution unchanged. 
Table \ref{tab:loss_sinu_td} lists the $L_2$ normalised errors and total loss for the time-dependent study and shows that the PINN model can capture the transient 
flow efficiently. The results at t= 0.5 are of similar order to those obtained for steady flow at $Re=100$ as presented in Table \ref{tab:loss_sinu}.

\subsection{Limitations and future applicability}

Despite the advantages of PINN, there are certain limitations that need to be addressed. A primary challenge lies in optimization of the hyperparameters, which can influence the model's performance. Secondly, the training time for PINN is often longer than traditional CFD simulations, especially for cases involving complex geometries, or high Reynolds numbers. This is due to the fact that standard PINNs use global neural network approximators, which tend to smooth out localized structures such as small scale vortices. Resolving such features typically requires either a very large network or a dense set of collocation points \cite{Nabian_2021} or a large number of iterations, all of which increase the computational cost.

Our results show that the PINN can effectively solve the benchmark problems at low or moderate Reynolds number. However, the predictions of the PINN model become less accurate at large $Re$ values. This can be inferred from Figure \ref{fig:lid_driven_uvp_all} and Figure \ref{fig:ghia_comp}, the PINN predicted flow patterns and velocity profiles for the lid-driven cavity flow at $Re=1000$ show relatively larger deviation from the CFD results. The $L_2$ normalized errors also increase with $Re$ as shown in Table \ref{tab:loss_lid}. 
The gradient-based training fails for these cases because the network gets stuck in local minima due to the coupled losses from multiple governing equations. The derivative terms in the Loss functions which are computed using AD further complicate the loss landscape especially for high-order derivatives. 

There have been significant efforts to improve the accuracy of PINN at high Reynolds numbers. Several literature \cite{karniadakis2021physics, cuomo2022scientific, en16052343, CHIU2022114909,wang2022and, jagtap2020locally,wang2020understanding} have investigated these limitations and proposed different strategies to overcome these issues. 
Control-volume based PINN (CV-PINN) \cite{10.1063/5.0256470} reformulate the continuity loss in integral form over the control volumes. Since the loss function is in form of algebraic equations, the complexity of the loss landscape due to the derivative terms is alleviated. 
Coupled automatic-numerical differentiation (CAN-PINN) \cite{CHIU2022114909} combines AD with Numerical Differentiation (ND) to reduce computational cost and stabilize gradient evaluation particularly useful for stiff viscous terms. 
To escape the local minima, re-initialization techniques \cite{Jongmok_reinitialization} have been proposed where the training parameters are periodically modulated in such a way that the regions in the collocation points where loss stagnates gets a random value. 
Several works have focused on integrating traditional numerical discretization methods in PINNs to enhance their accuracy and efficiency. For instance, the Least-Square-based-finite-differece PINN (LSFD-PINN) \cite{XIAO202433} uses LSFD method to compute the derivatives in the loss function, thus reducing the computational cost. 
Recent development also explores the coupling of PINNs with data assimilation (DA) methods \cite{satyadharma2024assessing}.

\section{Conclusion}
\label{sec:conclusion}

In this work, we have used an unsupervised physics-informed neural network (PINN) to study three steady-state incompressible flow problems along with a transient case which are governed 
by 2D Navier-Stokes equations. Instead of using labeled data in the training of DNN, we have implemented the initial conditions and the boundary conditions in a hard manner. 
To improve the training convergence and stability, we incorporate a learning rate scheduler in our PINN framework, which dynamically adjusts the learning rate whenever the loss function fails to decrease for a certain number of epochs.
We have considered three scenarios: lid-driven cavity flow, flow passing a circular obstacle, and pipe flow with a sinusoidal boundary. PINN results are further compared with the CFD simulations obtained using COMSOL Multiphysics. It is observed that there is a reasonable agreement between the results of PINN and CFD simulations. For the lid-driven cavity, we also present the comparisons of velocity fields at the centre-line of the cavity with the results obtained from CFD simulations and by Ghia et al. \cite{GHIA1982387} for $Re=100$. However at higher $Re = 1000$, the PINN produced values deviate away from CFD for the lid-driven scenario. PINN has also effectively captured the flow patterns, viscous and pressure drag coefficients for a 2D 
circular obstacle at different values of $Re$. It happens that PINN results agree well with the CFD values, although the total drag coefficient deviates by  $\sim 6\%$ at higher $Re$ (40). We also analyze the steady flow in a pipe with a sinusoidal boundary at different $Re$ values and subsequently extend the framework to study the transient flow in the same geometry with time-dependent inlet pressure.
The $L_2$ normalized errors lie in the range $\mathcal{O}(10^{-4}) - \mathcal{O}(10^{-1})$ for these case studies. The implementation 
of hard boundary conditions has reduced the number of terms in the loss functions 
and provides better accuracy compared to the soft boundary implementation. It is worth mentioning that although 
mesh-free PINN has several advantages, the current framework still requires longer training than CFD simulations. To further decrease the training time and increase the accuracy, the hyperparameters should be optimized more carefully. In the future scope, we aim to implement this PINN framework to solve fluid dynamics and heat transfer problems with more unconventional geometry and boundary conditions.

\vspace{+1cm}

\appendix

\section{Appendix A: Training history for different loss terms}
\label{appendix1}

\begin{figure}[!htb]
    \centering
        \includegraphics[width=0.9\textwidth]{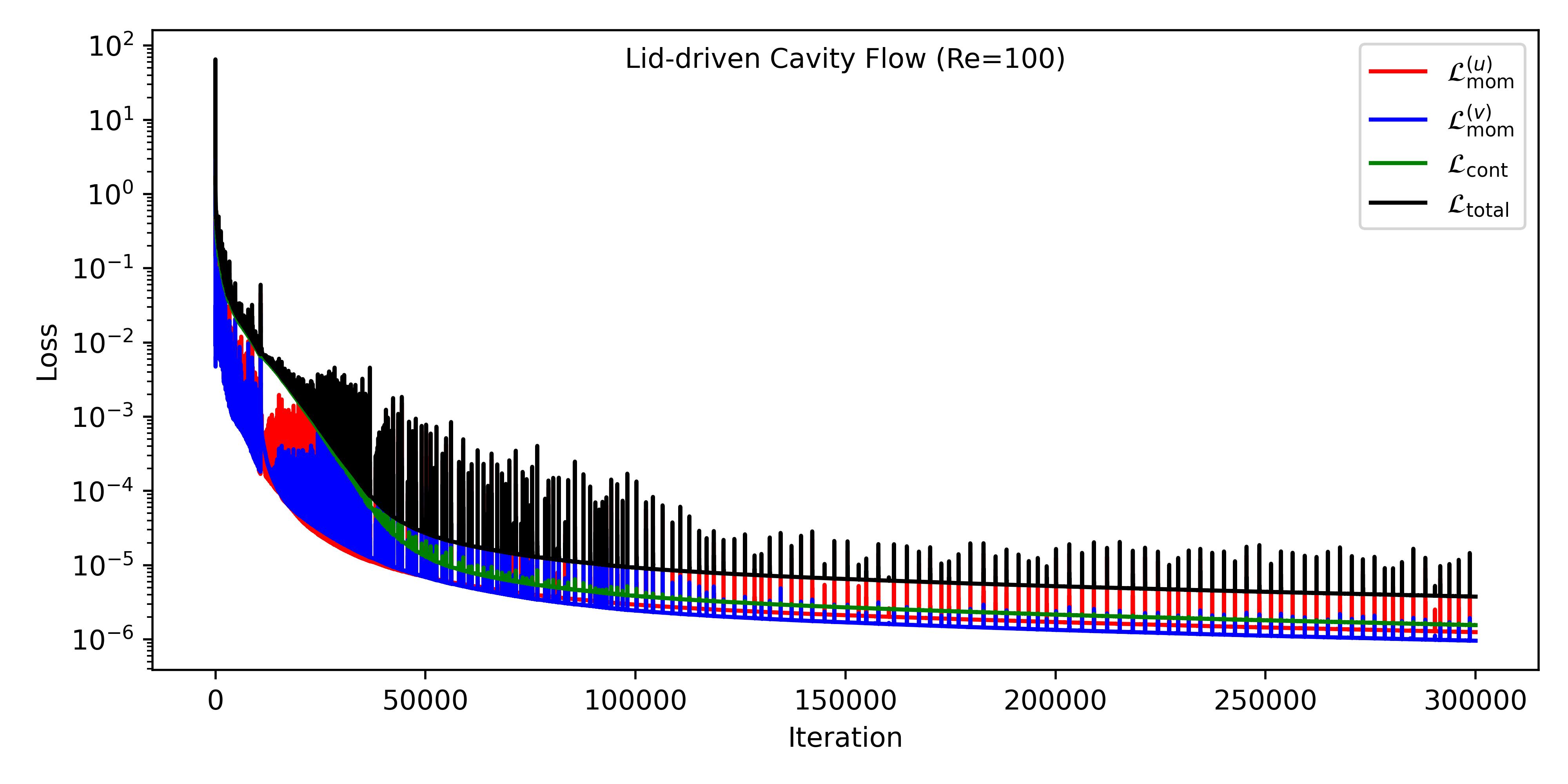}
      \caption{Evolution of loss terms and total loss over number of iterations for 2D lid-driven square cavity (steady-state flow) at Re = 100.}
              \label{fig:loss1}
\end{figure}

\begin{figure}[!htb]
    \centering
        \includegraphics[width=0.9\textwidth]{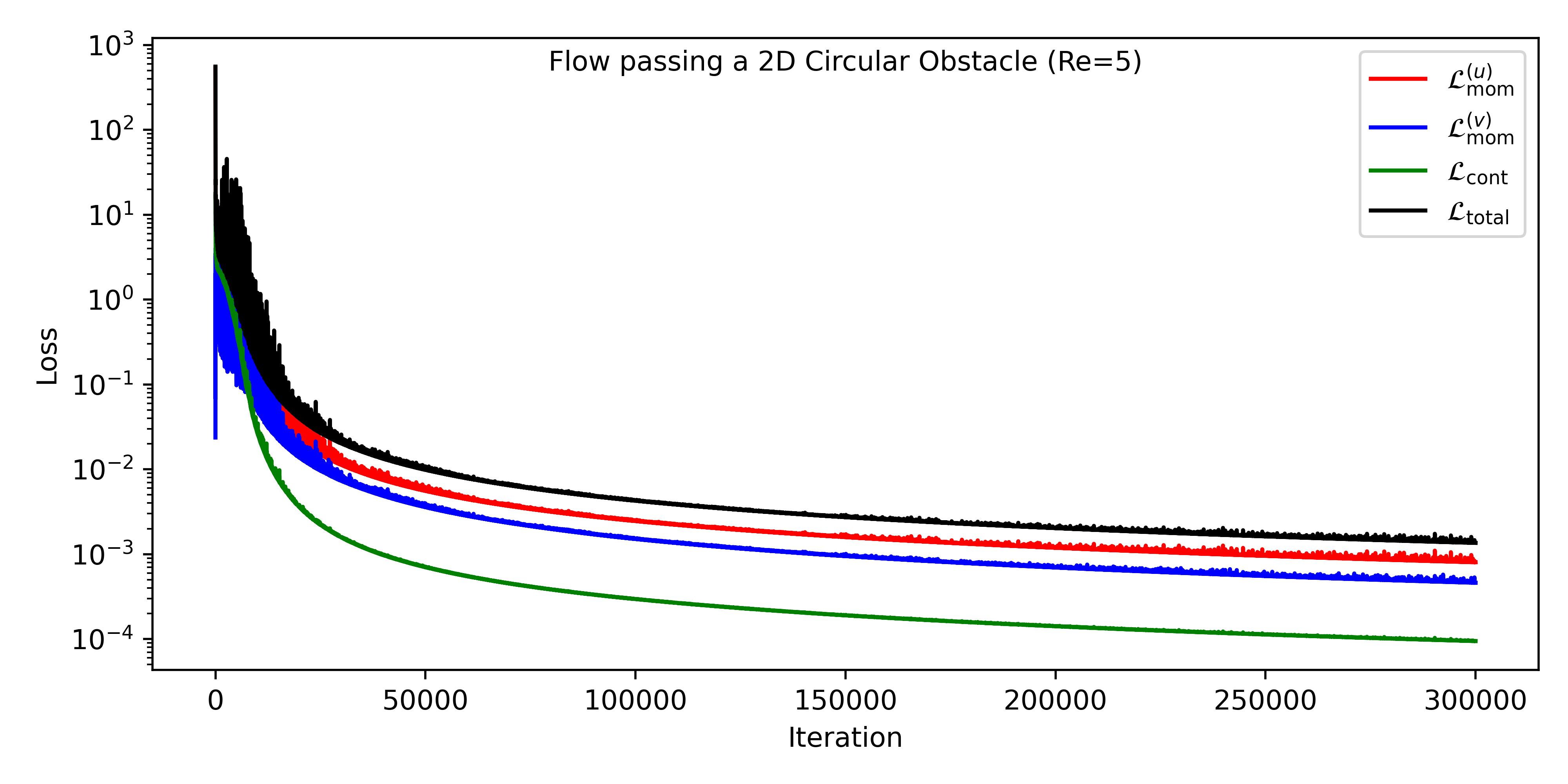}
      \caption{Evolution of loss terms and total loss over number of iterations for steady-state flow passing a 2D circular obstacle (Re=5).}
              \label{fig:loss2}
\end{figure}

\begin{figure}[!htb]
    \centering
                \includegraphics[width=\textwidth]{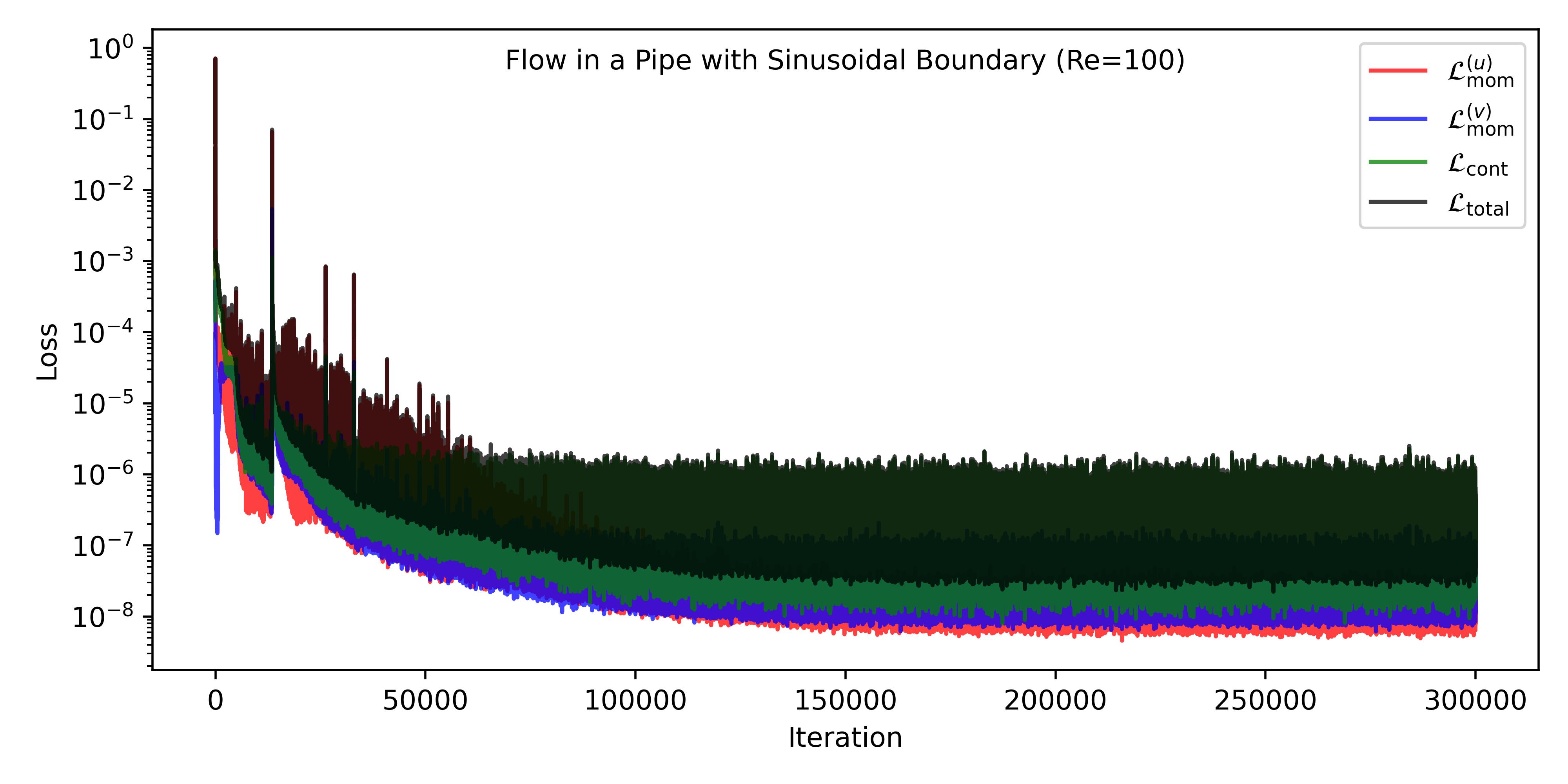}
      \caption{Evolution of loss terms and total loss over number of iterations for steady-state flow in a pipe with sinusoidal boundary (Re=100).}
              \label{fig:loss3}
\end{figure}

\begin{figure}[!htb]
    \centering
        \includegraphics[width=\textwidth]{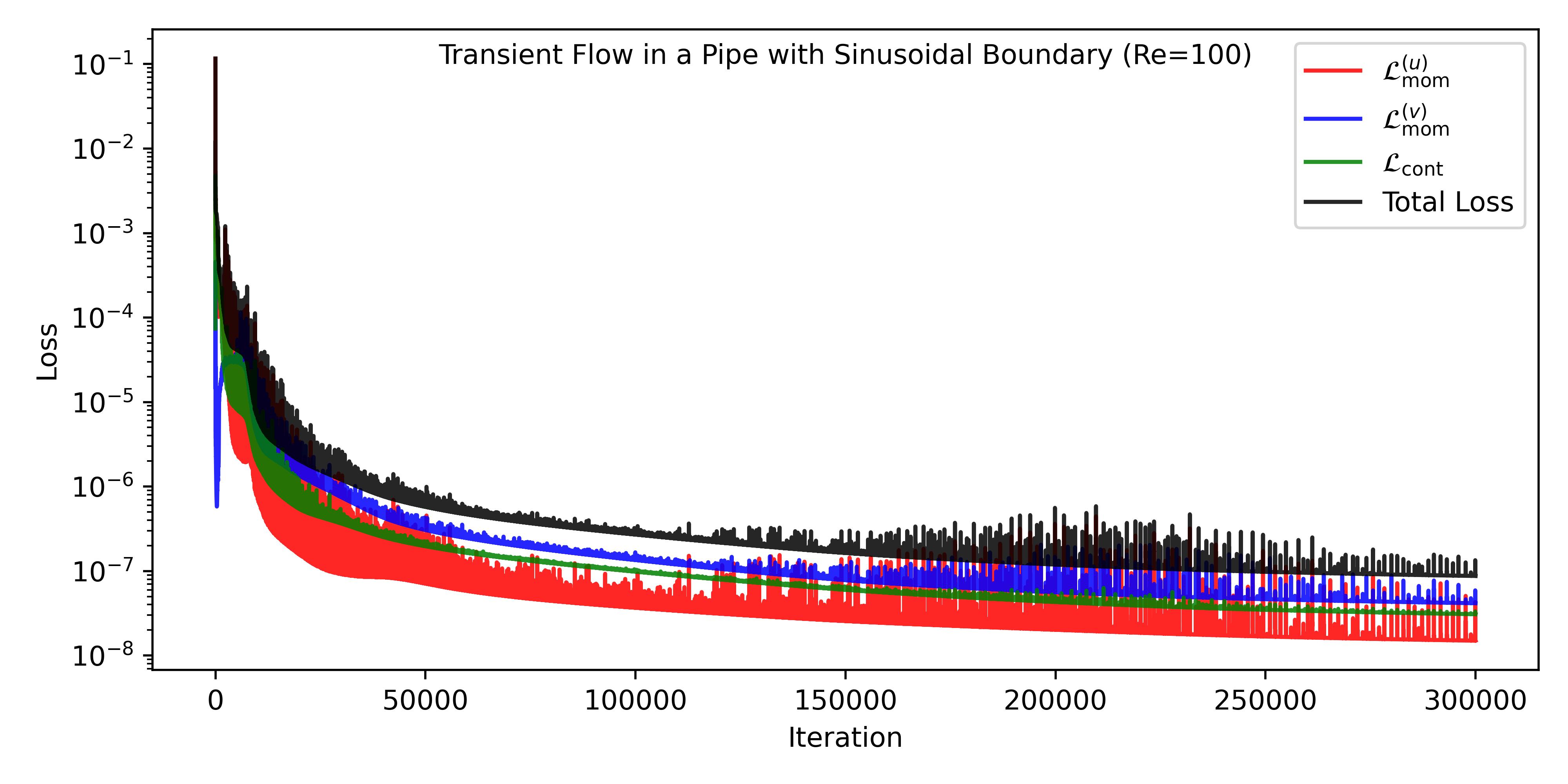}
      \caption{Evolution of loss terms and total loss over number of iterations for transient flow in a pipe with sinusoidal boundary (Re=100).}
              \label{fig:loss4}
\end{figure}

Figures \ref{fig:loss1} - \ref{fig:loss4} show the training history of our PINN framework for the four benchmark cases. We depict the evolution of losses due to the $x$-momentum ($\mathcal{L}_{\text{mom}}^{(u)}$), $y$-momentum ($\mathcal{L}_{\text{mom}}^{(v)}$) and the continuity equation ($\mathcal{L}_{\text{cont}}$) along with the total loss in each figure.

\newpage

\section{Appendix B: Hardware and Training time}

The computational costs for different flow scenarios are listed in Table \ref{tab:training_times} along with the corresponding hardware configurations. 
The total number of iterations in PINN for each case is $3 \times 10^5$.
The CFD simulations are performed on CPU-based machines and modern-day machine learning algorithms achieve faster performance on GPUs, except for the case of steady pipe flow in a sinusoidal boundary. This can be attributed to the fact that CPU cores are faster than GPU cores, therefore, for the training to benefit from large number of cores of a GPU, the network size or the batch size needs to be sufficiently large. However, the current PINN framework is still significantly more expensive than CFD as evident from Table \ref{tab:training_times}.

\begin{table}[!htb]
\centering
\renewcommand{\arraystretch}{1.3}
\begin{tabular}{|c||c||c|c|}
\hline
\textbf{Case} & {\textbf{Hardware}} & \multicolumn{2}{c|}{\textbf{Time}} \\
\cline{3-4}
 &  & CFD  & PINN\\
\hline
 Lid-driven cavity flow	&  Intel Xeon W5-2455X x24 64GB (CPU) &  2 s & $\sim$23.5 h \\
 \cline{2-4}
 (Re=100, steady flow)  				&  NVIDIA GeForce RTX 3080 10GB (GPU) & - & $\sim$3.5 h\\
\hline
\hline
 Circular obstacle		&  Intel Xeon W5-2455X x24 64GB (CPU) & 3 s   & $\sim$59.0 h \\
 \cline{2-4}
 (Re=5, steady flow)  				&  NVIDIA GeForce RTX 3080 10GB (GPU) & - & $\sim$7.5 h\\
 \hline
 \hline
   Steady pipe flow 	   	&  Intel Xeon W5-2455X x24 64GB (CPU) &  7 s & $\sim$2.0 h \\
 \cline{2-4}
 (Re=100)  				&  NVIDIA GeForce RTX 3080 10GB (GPU) & - & $\sim$ 3 h\\
 \hline
 \hline
    Transient pipe flow 	 &Intel Xeon W5-2455X x24 64GB (CPU)  & 340 s  & $\sim$ 53 h \\
 \cline{2-4}
	(Re=100)  					 &  NVIDIA GeForce RTX 3080 10GB (GPU) & - & $\sim$6.0 h\\
	 \hline
 \hline
\end{tabular}
        \caption{Training time for PINN and CFD in different test cases and hardware configurations. The total number of iterations in PINN for each case is $3 \times 10^5$.}
            \label{tab:training_times}
\end{table}

\section{Appendix C: Choosing the appropriate function for hard boundary conditions}
\label{appendixc}
The hard-boundary conditions are implemented using Equation \ref{eq:hard_bc}, where $A(x)$ can be chosen as any function that is zero at the boundary and non-zero elsewhere, and $B(x)$ can be chosen as the value of the velocity at the boundary.

For example, in the case of the lid-driven cavity, the boundaries lie at $x=0.0, 1.0$ and $y=0.0, 1.0$. (see Figure \ref{fig:lid_cavity_geo}).
For the function $u(x, y) = u_{NN}(x, y) A(x) + B(x)$, $A(x)$ must be zero at these locations and non-zero elsewhere. The most basic choice of $A(x)$ can be $A(x) = x (1-x) y (1-y)$. 
Next we introduce two parameters $\theta_1 $ and $ \theta_2$ as shown in Figure \ref{fig:lid_cavity_geo}, to incorporate the hard boundary conditions for horizontal velocity at the boundary walls. We see that at $\theta_1 = \theta_2 = 0$ and $y=0$ lines, we require $B(x) = 0$. This can be achieved by demanding $B(x)$ to be proportional to $y \theta_1 \theta_2$ \footnote{One could have chosen $B(x) \propto y x (1-x)$, but that would have made setting $B(x)=1$ at the top wall difficult as $x$ varies from 0 to 1.}.
Then at the top wall, $B(x)$ is proportional to $1 \times \frac{\pi}{2} \times \frac{\pi}{2} = \frac{\pi^2}{4}$. But we require $B(x)$ to be 1 at the top wall. Therefore, we can choose $B(x) = \frac{4}{\pi^2} y \theta_1 \theta_2$. This gives us the final form of $u(x, y)$ as given in Equation \ref{eqn:u1}. Similarly, we choose appropriate A(x) and B(x) in Equation \ref{eq:cir_hard}, Equation \ref{eq:sine_hard} and Equation \ref{eq:sine_hard_td} to implement the hard boundary conditions for the circular obstacle and sinusoidal pipe cases, respectively.

\section{Appendix D: Losses with and without hard boundary implementation}

\begin{figure}[!htb]
    \centering
        \includegraphics[width=0.79\textwidth]{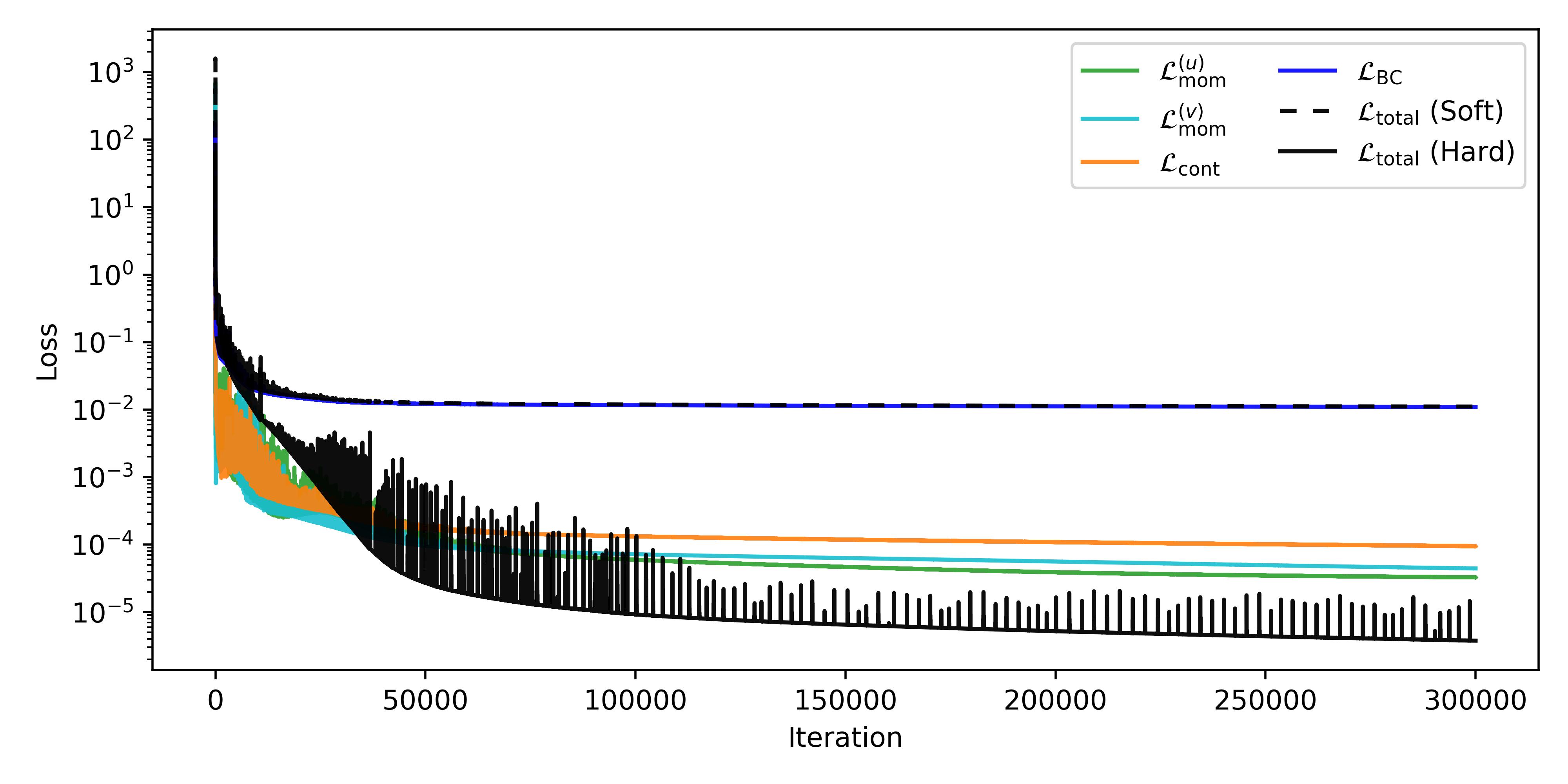}
    \caption{Evolution of different and total loss function without ``\texttt{hard}" boundary condition enforcement method for the lid-driven cavity with Re = 100. The black solid line represents the total loss for ``\texttt{hard}" BC implementation.}
    \label{fig:soft}
\end{figure}

The loss function after the implementation of ``\texttt{hard}" boundary condition 
is expressed as $\mathcal{L}_{Data-free-PINN}^{Hard-BC} = \mathcal{L_{\text{PDE}}} = 
   \mathcal{L}_{\text{mom}}^{(u)} + \mathcal{L}_{\text{mom}}^{(v)} + \mathcal{L_{\text{cont}}}$ and the individual loss terms are defined 
   in Equation \ref{eqn:loss_hard}. Without the ``\texttt{hard}" boundary enforcement, the loss function is modified to include additional contributions from boundary conditions (BC) and/or initial conditions (IC) depending on whether the flow is steady or transient. The total loss function is then given by 
      $ \mathcal{L}_{Data-free-PINN}^{Soft-BC} = \mathcal{L}_{\text{mom}}^{(u)} + \mathcal{L}_{\text{mom}}^{(v)} + \mathcal{L_{\text{cont}}}  + \mathcal{L}_{BC} + \mathcal{L}_{IC}$. 
Implementation of BC/ICs in the loss function in a soft manner 
does not always guarantee that these conditions will be satisfied exactly after PINN training and thus the PINN solutions may show a discrepancy. 
We demonstrate the effectiveness of enforcement of ``\texttt{hard}" BC by comparing the loss functions with the soft case in Figure \ref{fig:soft}. 
To illustrate the relative performances for these two approaches, we consider the lid-driven cavity problem with $Re =100$. Figure \ref{fig:soft} presents the evolution of losses associated with the $x$-momentum ($\mathcal{L}_{\text{mom}}^{(u)}$), $y$-momentum ($\mathcal{L}_{\text{mom}}^{(v)}$), the continuity equation ($\mathcal{L}_{\text{cont}}$), boundary conditions 
($\mathcal{L}_{\text{BC}}$) along with the total loss ($\mathcal{L}_{total}^{Soft-BC}$) for Soft-BC method. The individual loss terms along with total loss ($\mathcal{L}_{total}^{Hard-BC}$) for ``\texttt{hard}" BC method are already shown in Figure \ref{fig:loss1}. For comparison, the total loss 
$\mathcal{L}_{total}^{Hard-BC}$ is also indicated in Figure \ref{fig:soft} as a black solid line. It is evident that the boundary loss term has a dominant contribution in the Soft-BC enforcement method and the total loss is approximately three orders of magnitude larger than that in the Hard-BC case. This happens due to the fact that in the ``\texttt{hard}" boundary condition enforcement method, BCs are encoded in PINN and are automatically satisfied through Equation \ref{eq:hard_bc}.

\begin{table}[!htb]
\centering
\renewcommand{\arraystretch}{1.3}
\begin{tabular}{||c||c|c||c|c||}
\hline
\hline
\multirow{2}{*}{\textbf{Re}} & \multicolumn{2}{c||}{\textbf{$L_2$ normalized errors}} & \multicolumn{2}{c||}{\textbf{PINN}} \\
\cline{2-5}
 & $\epsilon_{L_2}^{(u)}$ & $\epsilon_{L_2}^{(v)}$ & Total Loss & No. of Iterations \\
\hline
100 (Soft-BC) & 0.08 & 0.11 & $1.1 \times 10^{-2}$ & 3,00{,}000 \\
\hline
100 (Hard-BC) & 0.04 & 0.08 & $3.77 \times 10^{-6}$ & 3,00{,}000 \\
\hline
\hline
\end{tabular}
\caption{Error metrics, total loss, and iteration count for the lid-driven cavity 
flow with and without ``\texttt{hard}" boundary implementation.}
\label{tab:loss_lid_soft}
\end{table}

\section*{Funding:} Work of UD was supported by Patna University minor research 
grant (01/RDC/RP/PU/Sanction).

\section*{Author Contributions:} All authors have equally contributed in this work and all authors have read and approved the final manuscript.

\noindent
\section*{Data availability statement:} The datasets are available from the authors: R.P. (email: ritik24@iisertvm.ac.in) and A.C. (email: arghya@iitp.ac.in) upon reasonable request. 

\bibliography{reference}

\end{document}